\documentclass{elsarticle}
\usepackage[utf8]{inputenc}
\usepackage{amsmath,amssymb,bm}
\usepackage{xfrac}
\usepackage{xcolor}
\usepackage[margin=1in]{geometry}
\usepackage{graphicx}
\graphicspath{{../figures/}}
\usepackage{tikz}
\usepackage{placeins}
\usepackage{tabularx}
\usepackage{booktabs}
\usepackage{lscape}
\usepackage[export]{adjustbox}
\newcolumntype{L}{>{\raggedright\let\newline\\\arraybackslash\hspace{0pt}}X}
\newcolumntype{Y}{>{\centering\arraybackslash}X}
\usepackage{siunitx}
\usepackage{algorithm, algpseudocode}
\algdef{SE}[DOWHILE]{Do}{doWhile}{\algorithmicdo}[1]{\algorithmicwhile\ #1}%
\usepackage{setspace}
\usepackage{caption}
\usepackage{subcaption}
\usepackage[colorlinks = true,
            linkcolor = blue,
            urlcolor  = blue,
            citecolor = blue,
            anchorcolor = blue,
            pdfauthor=author]{hyperref}
\usepackage[nameinlink]{cleveref}

\newcommand{\tensor}[1]{\bm{#1}}
\newcommand{\isotensor}[1]{\overline{\tensor{#1}}}
\newcommand{\iso}{\overline}
\newcommand{\tr}{\operatorname{tr}}
\renewcommand{\vec}[1]{\bm{#1}}        % vector valued function
\journal{}
\begin{document}
\begin{frontmatter}

\title{Efficient identification of myocardial material parameters and the stress-free reference configuration for patient-specific human heart models}
\author[GSRC,BTM]{Laura Marx}
\author[GSRC]{Justyna A. Niestrawska}
\author[GSRC]{Matthias A.~F. Gsell}
\author[GSRC]{Federica Caforio}
\author[GSRC,BTM]{Gernot Plank}
\author[GSRC]{Christoph M. Augustin\corref{cor1}}

\address[GSRC]{Gottfried Schatz Research Center for Cell Signaling, Metabolism and Aging - Division of Biophysics, Medical University of Graz, Graz, Austria}
\address[BTM]{BioTechMed-Graz, Graz, Austria}

\cortext[cor1]{
  Corresponding author at:
  Gottfried Schatz Research Center: Division of Biophysics,
  Medical University of Graz, Neue Stiftingtalstrasse 6/D04, 8010 Graz, Austria.
  E-mail address: (christoph.augustin@medunigraz.at (C.~M.~Augustin)
}

%\includegraphics[width=1.0\linewidth]{GraphicalAbstractPassiveFitting.pdf}

%Highlights are three to five (three to four for Cell Press articles) bullet points that help increase the discoverability
%of your article via search engines. These bullet points should capture the novel results of your research as well as new methods that were used during the study (if any).
%\item A novel model function-based fitting method to find personalized passive material parameters is introduced.
%\item The algorithm requires image data or previously generated meshes at only one time point and allows to fit
%to a patient-specific end-diastolic pressure-volume relation.
%\item An improved algorithm to find the stress-free reference configuration with a novel fail-safe feature is presented.
%\item The versatile workflow is applicable to a large variety of constitutive models and finite-element formulations.
%\item The method is robust, efficient and can be coupled to existing finite element software packages with relative ease.

\begin{abstract}
Image-based computational models of the heart represent a powerful tool to shed new light on the
mechanisms underlying physiological and pathological conditions in cardiac function and to improve diagnosis and therapy planning.
However, in order to enable the clinical translation of such models, it is crucial to develop
personalized models that are able to reproduce the physiological reality of a given patient.
There have been numerous contributions in experimental and computational biomechanics to
characterize the passive behavior of the myocardium.
However, most of these studies suffer from severe limitations and are not applicable to high-resolution geometries.
In this work, we present a novel methodology to perform an automated
identification of \emph{in vivo} properties of passive cardiac biomechanics.
The highly-efficient algorithm fits material parameters against the shape of a
patient-specific approximation of the end-diastolic pressure-volume relation (EDPVR).
Simultaneously, a stress-free reference configuration is generated,
where a novel fail-safe feature to improve convergence and robustness is implemented.
Only clinical image data or previously generated meshes at one time point
during diastole and one measured data point of the EDPVR are required as an input.
The proposed method can be straightforwardly coupled to existing finite element (FE) software packages
and is applicable to different constitutive laws and FE formulations.
Sensitivity analysis demonstrates that the algorithm is robust with respect to initial input parameters.
\end{abstract}

\end{frontmatter}

%journals: CMAME, Appl Math Model,

\section{Introduction}
Computational models of the heart are widely recognized as a powerful tool for the quantitative analysis of cardiac function. Their ability to explore mechanistic relationships of all variables of interest at high spatio-temporal resolution has turned computational models into an important, if not indispensable, adjunct in any basic research study. More recently though, driven by advances in medical imaging and simulation technologies, a translational trend has emerged that is geared towards turning cardiac modeling from a research tool into a clinical modality for diagnosis, stratification, and therapy planning~\cite{krishnamurthy2013patient, sermesant2012patient,Niederer2019computational}.

However, unlike in basic research applications where a representative ``one heart fits all'' model is suitable to investigate generic mechanisms, in clinical scenarios
%where modeling is used to inform clinical decisions,
the use of patient-specific models is of fundamental importance.
Computational models must be built to represent the physiological reality of a given patient to provide a basis for therapeutic decisions for the specific patient, not for a representative average patient.

This process of minimizing the difference between the predictions of a computational model and clinical observations, referred to as ``model personalization'' or ``digital twinning'', is challenging to achieve~\cite{Niederer2020}.
Cardiac function emerges as a complex interplay between different physics --- electrophysiology, mechanics, and hemodynamics.
While accurate mechanistic computational models of each of these physical components exist and anatimically accurate multiphysics models of the heart, e.g.,~\cite{Augustin2016Patient,Baillargeon2014,Gurev2015,Karabelas2018,Quarteroni2017,Santiago2018,Sugiura2012Multiscale}, are considered to be the state-of-the-art,
they comprise a larger number of parameters, most of which cannot be observed \emph{in vivo} or not at all.
Thus, model parameters have to be inferred indirectly from clinical observable quantities which are, in general, afflicted with a significant degree of observational and residual uncertainties~\cite{mirams2016uncertainty, Mirams2020}.

An important integral model component in modeling cardiac function are constitutive relations which characterize the passive biomechanical behavior of myocardial tissue. The development and parameterization of these models is an active area of research. As microstructural material components along local material axes determine the orthotropic behavior of the myocardium, recent experimental research has focused on multiaxial mechanical testing and microscopical investigations of the underlying structure \cite{Dokos2002Shear, Sommer2015biomechanical}. These experiments are usually performed under artificial \emph{ex vivo} conditions using excised tissue specimen from specific locations within the myocardium. While these experiments are an indispensable source for informing modeling, their inherent underlying limitations and uncertainties should be kept in mind when incorporating such data into models. The tissue is usually excised during autopsies within 12h of death and then stored in a suitable solution until testing. The testing should ideally take place immediately, but, in practice, may be postponed for weeks, although it has been shown for collageneous tissues that changes in microstructure and mechanics occur in as little as 48h after removal from the body \cite{Hemmasizadeh2012}. Available material parameters are usually fitted to a normalized stress-strain curve obtained from multiple patients and used to represent the overall passive mechanical behavior of the ensemble.
It is known though that tissue properties vary to a significant extent between individuals and throughout the myocardium. Additionally, even if orthotropic, microstructurally-based models are utilized, the underlying orthotropic fiber and sheet arrangement in the myocardium cannot be determined with certainty for individuals. To achieve translation it is therefore key to develop a methodology that facilitates the efficient, robust and, ideally, automated identification of model parameters from clinical observations \emph{in vivo}~\cite{Augustin2016Patient}.

A method towards \emph{in vivo} passive parameter estimation was proposed by \citet{Augenstein2005Method}. They utilized an \emph{ex vivo} experimental method based on the combination of tagged magnetic resonance imaging (MRI) and simultaneous pressure recordings to estimate passive material properties in arrested pig hearts. The fiber architecture was incorporated from diffusion tensor MRI and the findings were validated against a silicon gel phantom with known material properties, proving that cardiac MRI can be used to extract meaningful material properties. Further work on isolated hearts combined with FE analysis was then conducted by e.g. \citet{Nair2007Optimizing}, where an \emph{ex vivo} rabbit left ventricular (LV) model was used for strain matching.

To take a step towards the translation of passive material identification into the clinics, \emph{in vivo} MRI combined with \emph{ex vivo} diffusion tensor MRI was carried out in heart failure patients \cite{Wang2009Modelling} and, using a sequential quadratic programming optimization technique \cite{boggs1995sequential},  passive material parameters were identified \cite{Wang2013Changes}. These studies used either transversely isotropic laws such as the Guccione law \cite{Guccione1991Passive} or orthotropic laws such as the Holzapfel-Ogden (HO) model \cite{Holzapfel2009Constitutive}. Further, initial parameter sets fitted to \emph{ex vivo} experimental data (e.g. \cite{Gultekin2016}) are taken and either only the isotropic parameter is varied \cite{Wang2013Changes,Asner2016estimation,Finsberg2018efficient} or two scaling parameters \cite{Nasopoulou2017improved,krishnamurthy2013patient,sack2018construction,nordsletten2011coupling} are introduced to preserve the overall orthotropy of the material parameters. Additionally, material parameters of these laws are often correlated \cite{Remme2004Development}, and hence the unique identification poses a non-trivial problem.

Early studies attempted to identify material parameters based on non-invasive imaging of strain via MRI \cite{Nash2000Computational,Young2009computational}.
However, \emph{in vivo} MRI strain measurements have major caveats as they
vary among methods, modalities, and software version and mostly lack
proper validation~\cite{Amzulescu2019, Reichek2017}.
In addition, MRI strains are not usually acquired during routine clinical exams, and hence pressure-volume (PV) relations are more commonly used to estimate material parameters. There are two common fitting targets used in literature: fitting to displacement curves obtained from several frames during diastole or approximation of the empirical EDPVR by a power law as proposed by~\citet{Klotz2007Computational}.

Inverse estimation of patient-specific parameters is now widely used, where parameters of a forward problem are tuned to match PV-curves and motion fields available from clinical imaging~\cite{Gao2015parameter}, a method which was first used by~\citet{Ghista1980Cardiac}.
\citet{Xi2011Myocardial,Xi2013estimation} utilized a reduced-order unscented Kalman filter \cite{Moireau2011reduced} and later 3D tagged MRI to simulate unloading and estimate parameters in a reformulated Guccione law to overcome the problem of non-unique constitutive parameters. They used an early diastolic frame as the reference configuration and 29 frames per cycle for the fitting. \citet{Asner2016estimation} jointly estimated the reference state and passive parameters using deflation, considering a parameter sweep consisting of 25 simulations to assess the parameter space and find the minimum of the objective function. \citet{Gao2015parameter} used the first frame of early diastole as the unloaded reference configuration and assumed a population-based end-diastolic pressure, $p_\mathrm{ed}$, of \SI{8}{\mmHg} as no pressure recordings were available. They proposed a multi-step non-linear least-squared optimization procedure for the inverse estimation of all parameters of the HO law. The performance of multiple constitutive laws were compared by~\citet{Hadjicharalambous2015analysis}, who used 3D tagged MRI and a parameter sweep. The initial parameter set was chosen to match the Klotz relation. They found that the HO law provided the best balance between identifiability and model performance.  Recently,~\citet{Nasopoulou2017improved} used 2D tagged MRI to improve the identifiability of parameters of the Guccione law by combining deformation and energy analysis to uniquely constrain the parameters. In more detail, the exponential scaling factor $\alpha$ was estimated through the minimization of an energy-based cost function. Then, a mechanical simulation was performed in a second step to optimize the linear isotropic parameter with a cost function based on differences in displacements. For this, the first frame and the two last frames of diastole were used to be compared to simulations of passive inflation. \citet{Finsberg2018estimating,Finsberg2018efficient} used the backward displacement method to find the unloaded configuration and fitted one parameter of a reduced HO law with an initial guess of other parameters similar to~\citet{Asner2016estimation}.

The empirical EDPVR as proposed by Klotz et al. was utilized as a fitting target by, e.g.~\citet{nordsletten2011fluid}, to estimate the unloaded geometry of human hearts. Simulations of passive inflation were conducted and compared to the Klotz relation, using least-square constraints on volume and pressure. In 2013,~\citet{krishnamurthy2013patient} estimated the unloaded bi-ventricular geometry using a backward displacement method. They scaled parameters of a reduced HO law to fit the empirical Klotz curve by adjusting them manually and varying exponential parameters within a range of \SI{15}{\%} of their initial values taken from experiments performed on canine tissue \cite{Dokos2002Shear}. In 2018,~\citet{Palit2018In} provided a summary of parameters of the HO law available in literature and examined the influence of $p_\mathrm{ed}$, fiber orientations and geometry on the estimated parameters. They used an objective function which minimized the differences between the simulated and real LV cavity volume and used the Klotz curve as a fitting target albeit they had no pressure recordings and assumed $p_\mathrm{ed}$ to be \SI{10}{\mmHg}. The initial parameters were estimated by a Latin hypercube sampling (LHS) which generated 50 initial data sets.
\citet{sack2018construction} used a similar approach, minimizing the error with respect to both pressure and volume. Initial parameters were fitted to data from \cite{Sommer2015biomechanical} and scaled with one scaling parameter for the linear terms and one for the exponential terms to match the Klotz relation. For the passive filling calibration, they defined an objective function as the difference in pressure values along the PV-curve combined with a single measure of end diastolic volume, $V_\mathrm{ed}$.

In this paper, we present a novel model function-based fitting method (MFF) to find personalized material
parameters for passive mechanical modeling of the heart.
Only image data at one time point to create the anatomical model and one measured data point of the EDPVR are required as an input.
The approach simultaneously personalizes passive material parameters and generates a stress-free reference
configuration which is crucial for the image-based modeling of biomechanical LV diastolic function.
In this regard, we also present an improved backward displacement algorithm where robustness is enhanced by a novel fail-safe feature.
The MFF method was tested on a cohort of 19 LV meshes~\cite{Marx2020}. Excellent agreement
with the empirical Klotz EDPVR was obtained for all cases under study.
As the determination of a stress-free reference configuration and the identification of parameters is carried out
simultaneously, the method is highly efficient, requiring only a small number of forward simulations
($\leq 10$ for all cases). Hence, computational costs were only about 2 to 3 times the cost of a standard passive inflation experiment.
The versatile workflow is applicable to a large variety of constitutive models, model functions, and FE formulations
and can be coupled to established FE software packages with relative ease.
A thorough sensitivity analysis demonstrates the robustness of the method with regard to input uncertainty.
%genetic algorithms \cite{Nair2007Optimizing,Sun2009computationally, Mojsejenko2015estimating} cover whole parameter space. extensive number functional evaluations, computationally expensive. derivative free method.\\

%%%%%%%%%%%%%%%%%%%%%%%%%%%%%%%%%%%%%%%%%%%%%%%%%%%%%%%%%%%%%%%%%%%%%%%%%%%%%%%%%%%%%%%%%%%%%%%%%%%
%not sure:

%\citet{aguado2011patient} method for estimating the unloaded geometry based on the multiplicative decomposition of the deformation gradient tensor (as it is done in growth modelling). The objective of the iterative estimation scheme is to find the unloaded reference geometry that minimizes the difference between the measured end-diastolic geometry and the computed geometry when the unloaded model is inflated to the measured end-diastolic pressures.

 %\cite{nikou2016computational}   Multi-step optimization method (cost functional based on strain data and PV data to estimate passive parameters in Guccione and HO law. They used the optimization algorithm proposed by \cite{Gao2015parameter}

 %%%%%%%%%%%%%%%%%%%%%%%%%%%%%%%%%%%%%%%%%%%%%%%%%%%%%%%%%%%%%%%%%%%%%%%%%%%%%%%%%%%%%%%%%%%%%%%%%%%%%%

\section{Methods}
\subsection{Patient Data}
\label{sec:patientdata}
Clinical data from the CARDIOPROOF study (NCT02591940), a recent clinical trial, was available, which includes data of $N_\mathrm{AS} = 12$ aortic valve stenosis (AS) patients and  $N_\mathrm{CoA} = 7$ aortic coarctation (CoA) patients.
As indicator for treatment, valve area and/or systolic pressure drop across the valve was taken into account for AS cases, whereas for CoA cases treatment indicators included an echocardiographically measured peak systolic pressure gradient across the stenotic region greater than \SI{20}{\mmHg} (\SI{2.66}{\kPa}) and/or arterial hypertension. The institutional Research Ethics Committee approved the study following the ethical guidelines of the 1975 Declaration of Helsinki. From the participants' guardians written informed consent was attained. A detailed description of the data acquisition process and clinical protocols used in this study were reported in~\cite{Fernandes2017Beyond}. Pressure measurement in the LV, namely invasive catheterization, was routinely acquired in patients suffering from CoA only, making measured values of $p_\mathrm{ed}^\mathrm{dat}$ solely available for this group. For AS cases $p_\mathrm{ed}^\mathrm{dat}$ was determined empirically by statistically analyzing a reference data pool of $N=290$ patient cases treated for AS. For more details on the reference data pool see  the supplementary material of~\cite{Marx2020}.

\subsection{Stress-free reference configuration}
\begin{figure}\centering
  \begin{tikzpicture}[remember picture]
  \node[anchor=north west, inner sep=0pt] (structure) at (0.2,0)
    {\includegraphics[width=125mm]{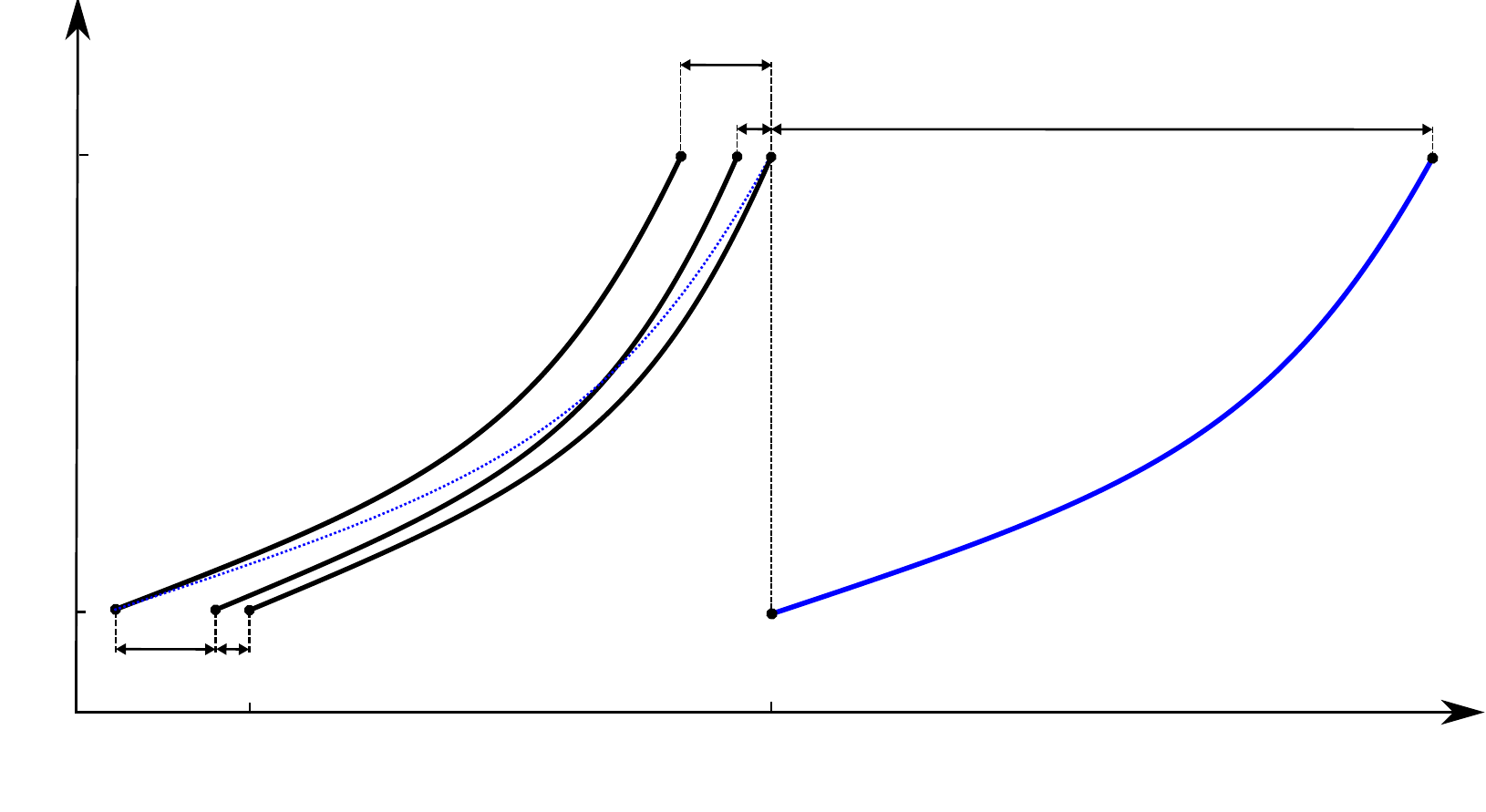}};
  \node at (11.2, -3.80) {\scriptsize $\text{it}_1$: $\bm{\phi}(\vec{X}^1)$};
  \node at (3.5, -3.80) {\scriptsize $\text{it}_2$};
  \node at (5.7, -2.30) {\scriptsize $\text{it}_3$};
  \node at (5.35, -3.80) {\scriptsize $\text{it}_4$};
  \node at (9.5, -0.83) {\scriptsize $\bm{R}^1=\bm{x}^{1}-\bm{x}^{\mathrm{dat}}$};
  \node at (12.6, -1.20) {\scriptsize $\bm{x}^{1}$};
  \node at (6.35, -0.30) {\scriptsize $\bm{R}^2$};
  \node at (6.5, -0.83) {\scriptsize $\bm{R}^3$};
  \node at (1.6, -5.7) {\scriptsize $\bm{R}^2$};
  \node at (2.2, -5.7) {\scriptsize $\bm{R}^3$};
  \node at (0.65, -5.10) {\scriptsize 0};
  \node at (0.55, -1.25) {\scriptsize $p_\mathrm{ed}$};
  \node at (7.1, -6.25) {\scriptsize $\vec{X}^1 = \vec{x}^\mathrm{dat}$};
  \node at (2.3, -6.25) {\scriptsize $\vec{X}^\ast$};
  %\node at (6.7, -6.55) {\normalsize Cavity volume [\si{\mL}]};
  %\node[rotate=90] at (0.2, -3.25) {\normalsize Pressure [\si{\kPa}]};

  \end{tikzpicture}
  \caption{Schematic for Sellier's backward displacement method, see~\Cref{alg:sellier}.
           A forward inflation problem $\bm{\phi}$ (blue curve) is solved in the first iteration ($\text{it}_1)$,
           where a measured \emph{in vivo} pressure $p_\mathrm{ed}$ is applied
           to an interim reference configuration with coordinates $\vec{X}^1=\vec{x}^\mathrm{dat}$.
           This generates an updated configuration with coordinates $\vec{x}^1$.
           Subsequently, the per node displacement vector between the updated deformed configuration and the target
           \emph{in vivo} configuration ($\bm{R}^{1}$) is computed.
           Finally, the reference configuration is updated by subtracting this per node displacement vector (blue dashed line).
           This iterative procedure is repeated until a given error tolerance $\epsilon$ between
           the computed reference configuration and the given \emph{in vivo} configuration is reached.
           Here, the algorithm converges to the stress-free reference configuration $\vec{X}^\ast$ after four iterations.}
  \label{fig:unloading}
\end{figure}
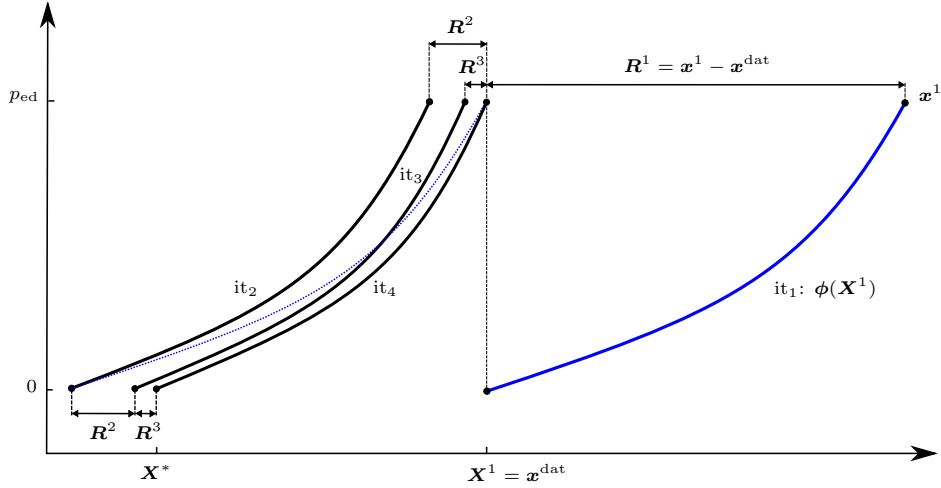

To find a stress-free reference configuration~\citet{Sellier2011Iterative} proposed an iterative, fixed-point method
for general elasto-static problems, see~\Cref{fig:unloading} for a schematic and~\Cref{alg:sellier}.
In each iteration $k$ a forward problem $\bm{\phi}(\vec{X}^k)$ is solved, where
a measured \emph{in vivo} pressure is applied to an interim reference configuration with coordinates $\vec{X}^k$.
This results in updated configurations with coordinates $\vec{x}^k$ and,
subsequently, the per node displacement vector between the updated deformed configuration and the target
\emph{in vivo} configuration $\bm{R}^k$ is computed.
Finally, the reference configuration is updated by subtracting this per node displacement vector.
This iterative procedure is repeated until a given error tolerance $\epsilon$ between
the computed reference configuration and the given \emph{in vivo} configuration is reached and
hence the stress-free reference configuration $\vec{X}^\ast$ is found.
\begin{algorithm}
\caption{Sellier's Inverse Method~\cite{Sellier2011Iterative}}%
\label{alg:sellier}
\begin{spacing}{1.2}
\begin{algorithmic}[1]
\State{initialize $\bm{X}^{1} = \bm{x}^{\mathrm{dat}}$; $k=0$}
\Do
\State{update counter, $k = k+1$}
\State{solve forward problem, $\bm{x}^{k}=\bm{\phi}\left(\bm{X}^{k}\right)$}
\State{calculate nodal error vector, $\bm{R}^{k}=\bm{x}^{k}-\bm{x}^{\mathrm{dat}}$}
\State{update reference vector, $\bm{X}^{k+1}=\bm{X}^{k}-\bm{R}^{k}$}
\doWhile{$\left\lVert\bm{R}^{k}\right\rVert \geq \epsilon$}
\State{stress-free reference configuration, $\bm{X}^{*}=\bm{X}^{k}$}
\end{algorithmic}
\end{spacing}
\end{algorithm}
\citet{Rausch2017Augmented} augmented this approach, see~\Cref{alg:aug_sellier}, based on Aitken's delta-squared process~\cite{aitken1950} that
increases the convergence rate and also significantly accelerates the method.
\begin{algorithm}
\caption{Augmented Sellier's Inverse Method~\cite{Rausch2017Augmented}}%
\label{alg:aug_sellier}
\begin{spacing}{1.2}
\begin{algorithmic}[1]
\State{initialize $\bm{X}^{1} = \bm{x}^{\mathrm{dat}}$; $k=0$; $\beta=1.0$}
\Do
\State{update counter, $k = k+1$}
\State{solve forward problem, $\bm{x}^{k}=\bm{\phi}\left(\bm{X}^{k}\right)$}
\State{calculate nodal error vector, $\bm{R}^{k}=\bm{x}^{k}-\bm{x}^{\mathrm{dat}}$}
\If{$k>1$}
\State{\(\displaystyle\text{update augmentation parameter},\ \beta = -\beta
       \frac{\bm{R}^{k-1}:\left[\bm{R}^{k}-\bm{R}^{k-1}\right]}{\left[\bm{R}^{k}-\bm{R}^{k-1}\right]:\left[\bm{R}^{k}-\bm{R}^{k-1}\right]}\)}
\EndIf
\State{update reference vector, $\bm{X}^{k+1}=\bm{X}^{k}-\beta \bm{R}^{k}$}
\doWhile{$\left\|\bm{R}^{k}\right\| \geq \epsilon$}
\State{stress-free reference configuration, $\bm{X}^{*}=\bm{X}^{k}$}
\end{algorithmic}
\end{spacing}
\end{algorithm}
Both fixed-point approaches are simple and versatile and can be coupled to existing FE software packages with relative ease and have thus been
applied to a number of biomechanical modeling problems,
e.g.,~\cite{Kallhovd2019,Karabelas2018,wittek2013vivo}.

However, for some of our LV models the augmented approach still diverged and unfavorable updates of the reference vector even resulted in a failure of the algorithm.
Since our parameter fitting approach described later in~\Cref{sec:model_function_fitting} is based on a robust unloading strategy for soft tissues, we further improved this iterative method by a fail-safe feature based
on an Armijo strategy~\cite{Armijo1966Minimization}, see~\Cref{alg:aug_sellier_armijo}.
Here, unfavorable search directions are damped and with this improvement the unloading algorithm was robust for all cases
and in most cases the procedure was sped up.

We successfully applied~\Cref{alg:aug_sellier_armijo} to all 19 LV models of the cohort from~\citet{Marx2020}, as well as meshes of bi-ventricular and four-chamber models from the cohort presented by~\citet{Strocchi2020publicly}.
%Let $\Lambda_{k} \subset\left\{1, \frac{1}{2}, \frac{1}{4}, \ldots, \lambda_{\min }\right\}$ denote a sequence such
%that
%\[
%T\left(x^{k}+\lambda_{k} \Delta x^{k}\right) \leq\left(1-\frac{1}{2} \lambda_{k}\right) T\left(x^{k}\right), \quad \lambda \in \Lambda_{k},
%\]
%holds and define an optimal damping factor via
%\[
%T\left(x^{k}+\lambda_{k} \Delta x^{k}\right)=\min_{\lambda \in \Lambda_{k}} T\left(x^{k}+\lambda \Delta x^{k}\right).https://www.overleaf.com/project/5ebc0e2882c75c000122175b
%\]

\begin{algorithm}
\caption{Augmented Sellier's Inverse Method with Armijo strategy}%
\label{alg:aug_sellier_armijo}
\begin{spacing}{1.2}
\begin{algorithmic}[1]
\State{initialize $\bm{X}^{1} = \bm{x}^{\mathrm{dat}}$; $k=0$; $\beta=1.0$}
\Do
\State{update counter, $k = k+1$}
\State{$\Lambda_{k} \subset\left\{1, \frac{1}{2}, \frac{1}{4}, \ldots, \lambda_{\min }\right\}$}
\For{$\lambda\in\Lambda_k$}
\State{solve forward problem, $\bm{x}^{\lambda}=\bm{\phi}\left(\bm{X}^{k}\right)$}
\State{calculate nodal error vector, $\bm{R}^{\lambda}=\bm{x}^{\lambda}-\bm{x}^{\mathrm{dat}}$}
\If{$k>1$}
\State{\(\displaystyle
\text{update augmentation parameter,}\
\beta = -\beta
       \frac{\bm{R}^{k-1}:\left[\bm{R}^{\lambda}-\bm{R}^{k-1}\right]}{\left[\bm{R}^{\lambda}-\bm{R}^{k-1}\right]:\left[\bm{R}^{\lambda}-\bm{R}^{k-1}\right]}
\)}
\EndIf
\State{update reference vector, $\bm{X}^{\lambda}=\bm{X}^{k}-\lambda\beta \bm{R}^{\lambda}$}
\State{\textbf{if} $\bm{R}^{\lambda}<\bm{R}^{k-1}$ \textbf{break}}
\EndFor
\State{\(\text{determine}\ \lambda^\ast \text{ such that }  \left\lVert\bm{R}^{\lambda^\ast}\right\rVert=\displaystyle\min_{\lambda \in \Lambda_{k}} \left\{\left\lVert\bm{R}^{\lambda}\right\rVert\right\} \)\vspace{0.5em}}
\State{update, $\bm{x}^{k}=\bm{x}^{\lambda^\ast}$, $\bm{R}^{k}=\bm{R}^{\lambda^\ast}$, $\bm{X}^{k+1}=\bm{X}^{\lambda^\ast}$}
\doWhile{$\left\lVert\bm{R}^{k}\right\rVert \geq \epsilon$}
\State{stress-free reference configuration, $\bm{X}^{*}=\bm{X}^{k}$}
\end{algorithmic}
\end{spacing}
\end{algorithm}

\subsection{Constitutive Fitting}
\subsubsection{Constitutive Models}
The myocardium is considered as a non-linear, hyperelastic, nearly incompressible, and orthotropic material
with a layered organization of myocytes and collagen fibers that is
characterized by a right-handed orthonormal set of basis vectors.
These basis vectors consist of the fiber axis $\vec{f}_0$, which coincides with
the orientation of the myocytes, the sheet axis $\vec{s}_0$ and
the sheet-normal axis $\vec{n}_0$.
To enforce the condition of a nearly incompressible material, the strain energy function $\Psi$,
is split into a volumetric $U(J)$ and an isochoric part $\Psi_\mathrm{isc}(\tensor{C})$:
\begin{equation}
    \label{eq:const-strain}
    \Psi(\tensor{C}) = U(J) + \Psi_\mathrm{isc}(\tensor{C}).
\end{equation}
In this relation, $\tensor{C}=\tensor{F}^\top\tensor{F}$ is the right Cauchy--Green deformation tensor, with $\tensor{F}$ the deformation gradient.
$U(J)$ is composed of the bulk modulus $\kappa \gg \SI{0}{\kPa}$ and a penalty term related to the Jacobian of the deformation gradient $J=\det(\tensor{F})$
and we choose
\begin{equation}
   U(J) = \frac{\kappa}{2}\ln(J)^2
\end{equation}
for all considered material models.
One of the simplest constitutive laws to model rubber-like materials is the isotropic \emph{Demiray model}~\cite{Demiray1972Note}
\begin{equation} \label{eq:Demiray}
  \Psi(\tensor{C})= U(J)
  + \frac{a}{2b}\left\{
  \exp\left[b(\iso{I}_1-3)\right]-1\right\},
\end{equation}
with the invariant $\iso{I}_1=\tr(\isotensor{C})$ and parameters $a>\SI{0}{\kPa}$ and $b>0$.
Here, $\isotensor{C}:=J^{-2/3}\tensor{C}$ is the isochoric part of the right
Cauchy--Green tensor resulting from the multiplicative split proposed by~\citet{Flory1961} to
model the nearly incompressible behavior of biological tissue.

\citet{Guccione1995Finite}, \citet{Usyk2000Effect}, \citet{Costa2001Modelling} and others introduced
\emph{single Fung-type exponential models} for the myocardium that can be generalized to
\begin{equation}\label{eq:UsykMaterial}
  \Psi(\tensor{C})
  = U(J)
  + \frac{a}{2}\left[\exp(\iso{Q})-1\right],
\end{equation}
where
\begin{equation} \label{eq:Usyk2000Q}
  \iso{Q} =  b_\mathrm{ff} \iso E_\mathrm{ff}^2
          +  b_\mathrm{ss} \iso E_\mathrm{ss}^2
          +  b_\mathrm{nn} \iso E_\mathrm{nn}^2
          + 2b_\mathrm{fs} \iso E_\mathrm{fs}^2
          + 2b_\mathrm{fn} \iso E_\mathrm{fn}^2
          + 2b_\mathrm{ns} \iso E_\mathrm{ns}^2.
\end{equation}
Here,
\begin{align*}
    \iso E_{\mathrm{ff}}&=\vec{f}_0\cdot\isotensor{E}\vec{f}_0,\
    \iso E_{\mathrm{ss}}=\vec{s}_0\cdot\isotensor{E}\vec{s}_0,\
    \iso E_{\mathrm{nn}}=\vec{n}_0\cdot\isotensor{E}\vec{n}_0,\\
    \iso E_{\mathrm{fs}}&=\vec{f}_0\cdot\isotensor{E}\vec{s}_0,\
    \iso E_{\mathrm{fn}}=\vec{f}_0\cdot\isotensor{E}\vec{n}_0,\
    \iso E_{\mathrm{ns}}=\vec{n}_0\cdot\isotensor{E}\vec{s}_0
\end{align*}
are projections of the Green-Lagrange strain tensor $\isotensor{E}=\frac{1}{2}(\isotensor{C}-\tensor{I})$
and parameter $a>\SI{0}{\kPa}$ serves as a stress-like scaling.
The dimensionless parameters $b_\mathrm{ff},b_\mathrm{ss},b_\mathrm{nn}>0$
account for the mechanical behavior of the tissue along the fiber (f), sheet (s) and sheet-normal (n) direction,
and $b_\mathrm{fs},b_\mathrm{fn},b_\mathrm{ns}>0$ account for structural interactions.
For parameter values of the different single Fung-type exponential models considered, see~\Cref{tab:singleFungMaterialParameters}.

\citet{Holzapfel2009Constitutive} proposed a \emph{separated Fung-type exponential model}
which can be generalized to
\begin{align}
  \Psi(\tensor{C})
    &= U(J)
    + \frac{a}{2b}\left\{
      \exp\left[b(\iso{I}_1-3)\right]-1\right\}
    + \sum_{i=\mathrm{f,s,n}}\frac{a_i}{2b_i}\left\{\exp\left[b_i
      ({I}_{4i}-1)^2\right]-1\right\}\nonumber\\
    &+ \sum_{i=\mathrm{fs,fn,sn}}\frac{a_{ij}}{2b_{ij}}\left\{\exp\left[b_{ij}
      ({I}_{8ij})^2\right]-1\right\},
 \label{eq:HolzapfelOgdenMyocardialModel}
\end{align}
with parameters $a_{(\bullet)} \ge \SI{0}{\kPa}$ and $b_{(\bullet)} > 0$, invariant $\iso{I}_1=\tr(\isotensor{C})$, unimodular fourth-invariants
\[
{I}_{4\mathrm{f}}= \max\left(\vec{f}_0\cdot\tensor{C}\vec{f}_0,1\right),\quad
{I}_{4\mathrm{s}}= \max\left(\vec{s}_0\cdot\tensor{C}\vec{s}_0,1\right),\quad
{I}_{4\mathrm{n}}= \max\left(\vec{n}_0\cdot\tensor{C}\vec{n}_0,1\right),
\]
such that contributions of compressed fibers are excluded, and interaction-invariants
\[
{I}_{8\mathrm{fs}}= \vec{f}_0\cdot\tensor{C}\vec{s}_0,\quad
{I}_{8\mathrm{fn}}= \vec{f}_0\cdot\tensor{C}\vec{n}_0,\quad
{I}_{8\mathrm{sn}}= \vec{s}_0\cdot\tensor{C}\vec{n}_0.
\]
Note that for the anisotropic contributions in~\Cref{eq:HolzapfelOgdenMyocardialModel}
the deformation gradient $\tensor{F}$
and the right Cauchy--Green tensor $\tensor{C}$ remain unsplit to avoid nonphysical
results~\cite{helfenstein2010non,sansour2008physical}.

Comparing to experimental data, Holzapfel and
Ogden reduced the constitutive law with the full set of invariants~\eqref{eq:HolzapfelOgdenMyocardialModel} to a
simplified model with the parameters
$a, a_\mathrm{f}, a_\mathrm{s}, a_\mathrm{fs} > \SI{0}{\kPa}$ and
$b, b_\mathrm{f}, b_\mathrm{s}, b_\mathrm{fs} > 0$.

Further, several papers~\cite{Eriksson2013modeling, Gultekin2016, Sommer2015biomechanical}
deal with in-plane and out-of plane dispersion of collagen fibers along the
$\vec{f}_0$ and $\vec{s}_0$ direction.
According to~\citet{Eriksson2013modeling} this is modeled by modifying the unimodular
fourth-invariants to
\begin{equation}\label{eq:fiber_dispersion}
{I}^\ast_{4i}=\kappa_i\overline{I}_1+(1-3\kappa_i)I_{4i}, \quad i\in\mathrm{f,s}.
\end{equation}
Dispersion parameters have been identified previously by mechanical experiments on passive
cardiac tissue by~\citet{Sommer2015biomechanical} and are set to $\kappa_\mathrm{f}=0.08$
and $\kappa_\mathrm{s}=0.09$.

For parameter values of the considered separated Fung-type exponential models, also often referred to as \emph{HO-type models}, see~\Cref{tab:HolzapfelOgdenMaterialParameters}.

%------------------------------------------------------------------------------------------------
\begin{landscape}
\begin{table}[htbp] \small
\setlength{\extrarowheight}{3.0pt}
\centering
\begin{tabularx}{0.62\linewidth}{*{9}{c}Y}
\toprule
Model & Ref. & Property & \multicolumn{7}{l}{Parameters} \\
\cline{4-10}
 & & & $a$ {\scriptsize[\si{\kPa}]} &
$b_{\mathrm{ff}}$   & $b_{\mathrm{ss}}$ &
$b_{\mathrm{nn}}$   & $b_{\mathrm{fs}}$ &
$b_{\mathrm{fn}}$   & $b_{\mathrm{ns}}$ \\
\midrule
Guccione & \cite{Guccione1995Finite} & trans.-isotr. &
$0.876$ & $18.48$ & $3.58$   & $3.58$  & $3.58$ & $1.627$ &$1.627$ \\
Usyk & \cite{Sugiura2012Multiscale} & orthotropic &
  $0.88$ & $5.0$  & $6.0$ & $3.0$ &$10.0$  & $2.0$ & $2.0$ \\
\bottomrule
\end{tabularx}
\caption{Default material parameters for the considered single Fung-type exponential materials.
Column `Model' gives the short name for the constitutive law;
`Ref.' the reference for the parameter fitting;
`Property' the material property specified by the parameter set;
and in the following a list of the material parameters.
Note that only the scaling parameter $a$ has a unit (\si{\kPa}), all other
parameters are dimensionless.}
\label{tab:singleFungMaterialParameters}
\end{table}

\begin{table}[htbp]   \small
\resizebox{\linewidth}{!}{
\setlength\tabcolsep{4.5pt} % default value: 6pt
\setlength{\extrarowheight}{3.0pt}
\centering
\begin{tabularx}{1.12\linewidth}{*{16}{c}Y}
\toprule
Model & Ref. & Property & \multicolumn{13}{l}{Parameters} \\
\cline{4-17}
& & & $a$  {\scriptsize[\si{\kPa}]}        & $b$ &
$a_{\mathrm{f}}$  {\scriptsize[\si{\kPa}]}& $b_{\mathrm{f}}$    &
$a_{\mathrm{s}}$  {\scriptsize[\si{\kPa}]} & $b_{\mathrm{s}}$    &
$a_{\mathrm{n}}$  {\scriptsize[\si{\kPa}]} & $b_{\mathrm{n}}$    &
$a_{\mathrm{fs}}$ {\scriptsize[\si{\kPa}]} & $b_{\mathrm{fs}}$   &
$a_{\mathrm{fn}}$ {\scriptsize[\si{\kPa}]} & $b_{\mathrm{fn}}$   &
$a_{\mathrm{sn}}$ {\scriptsize[\si{\kPa}]} & $b_{\mathrm{sn}}$   \\
\midrule
General HO & \cite{Guan2019AIC} & orthotropic &
0.180 & 9.762 & 2.204 & 21.597 & 0.098 & 49.878 & 0.508 & 27.719 & 1.291 & 5.295 & 1.345 & 2.017 & 0.947 & 4.514 \\
Reduced HO & \cite{Guan2019AIC} & orthotropic &
0.809 & 7.474 & 1.911 & 22.063 & -- & -- & 0.227 & 34.802 & 0.547 & 5.691 & -- & -- & -- & --\\
Original HO & \cite{Holzapfel2009Constitutive} & orthotropic &
0.33 & 9.242 & 18.535 & 15.972 & 2.564 & 10.446 & -- & -- & 0.417 & 11.602 & -- & -- & -- & --\\
HO dispersion& \cite{Gultekin2016} & orthotropic &
0.4 & 6.55 & 3.05 & 29.05 & 1.25 & 36.65 & -- & -- & 0.15 & 6.28 & -- & -- & -- & --\\
One fiber HO & \cite{Holzapfel2009Constitutive} & trans.-isotr. &
0.809 & 7.474 & 1.911 & 22.063 & -- & -- & -- & -- & -- & -- & -- & -- & -- & -- \\
Demiray & \cite{Demiray1972Note} & isotropic &
$1.0$ & 6.5 & -- & -- & -- & -- & -- & -- & -- & -- & -- & -- & -- & -- \\
\bottomrule
\end{tabularx}
}
\caption{Default material parameters for the considered Holzapfel--Ogden (HO) type materials.
Column `Model' gives the short name for the constitutive law;
`Ref.' the reference for the parameter fitting;
`Property' the material property specified by the parameter set;
and in the following a list of the material parameters.
Note that parameters $a_{(\bullet)}$ have a unit (\si{\kPa}), while
parameters $b_{(\bullet)}$ are dimensionless.
Model ``HO dispersion'' includes dispersion of collagen fibers based on~\eqref{eq:fiber_dispersion}.}
\label{tab:HolzapfelOgdenMaterialParameters}
\end{table}
\end{landscape}
%------------------------------------------------------------------------------------------------
\subsubsection{Fitting with model function}\label{sec:model_function_fitting}
Due to limited availability of clinical data representing the EDPVR, the passive biomechanical material model is fitted to the empirical Klotz EDPVR estimated from a single measured PV-pair~\cite{Klotz2007Computational}.
A more detailed description about the method proposed by ~\citet{Klotz2007Computational} can be found in~\ref{app:klotz}.

Fitting is done by adjusting material parameters of the constitutive relation during the unloading procedure (\Cref{alg:aug_sellier_armijo})
such that the cavity volume of the stress-free reference configuration $\vec{X}^\ast$ matches $V_0^\mathrm{klotz}$, see~\Cref{eq:const-v0},
and such that the simulated PV-curve matches the EDPVR as predicted by \Cref{eq:const-powerlaw}.
In particular, the parameters are varied in each unloading step as follows:
consider the function
\begin{align}
  \Phi(x, x_0)=\frac{a}{2b}\left\{
  \exp\left[b\left(\frac{x-x_0}{x_0}\right)\right]-1\right\}, \label{eq:model_function}
\end{align}
with parameters $a$ and $b$.
We use a Levenberg--Marquardt least-squares algorithm to fit the model function $\Phi(x, x_0)$
to the Klotz relation, where $x \in \{V^\mathrm{klotz}_i(p_i): i=1,...,n_\mathrm{ls}\}$ are the volumes as predicted by the Klotz power law~\eqref{eq:const-powerlaw} and
$x_0$ is the predicted stress-free volume $V_0^\mathrm{klotz}$~\eqref{eq:const-v0}. Here, $n_\mathrm{ls}$ is the number of loading steps and $p_i$ are equidistant pressures
in the interval $[0, p_\mathrm{ed}]$.
After convergence of the fitting method we get the parameter set
$[a_\mathrm{fit}^\mathrm{klotz}, b_\mathrm{fit}^\mathrm{klotz}]$.
Both fitting parameters are always strictly positive, due to the
convexity of the Klotz relation and the design of the model function~\eqref{eq:model_function}.
In a second step the procedure is repeated for the simulated PV-curve in the current step $k$ of the unloading algorithm with
$x\in \{V^\mathrm{sim}_i(p_i): i=1,...,n_\mathrm{ls}\}$
being the volumes at the different loading points and
$x_0$ the cavitary volume of the current reference configuration $\bm{X}^k$, to obtain the parameter set
$[a_\mathrm{fit}^\mathrm{sim}, b_\mathrm{fit}^\mathrm{sim}]$.
Note that $a_\mathrm{fit}^\mathrm{sim}>0$ holds, due to the fact that an increase in pressure leads to an increase in volume,
and $b_\mathrm{fit}^\mathrm{sim}\neq 0$, due to the design of the model function~\eqref{eq:model_function}.
We compute the scalings
\begin{equation} \label{eq:scalings}
  a^\mathrm{scale}=\frac{a_\mathrm{fit}^\mathrm{klotz}}{a_\mathrm{fit}^\mathrm{sim}},\quad
  b^\mathrm{scale}=\frac{b_\mathrm{fit}^\mathrm{klotz}}{|b_\mathrm{fit}^\mathrm{sim}|}.
\end{equation}
Here, the absolute value of $b_\mathrm{fit}^\mathrm{sim}$ is chosen, as this fitting parameter might be
negative in rare circumstances when the PV-curve for the
given model and material parameters is concave. Also, for the unlikely case of a linear PV-curve,
$b_\mathrm{fit}^\mathrm{sim}$ is close to 0, leading to a very large scaling parameter $b^\mathrm{scale}$.
Hence, to avoid large jumps in the material parameters, we define the bounding interval $\mathcal{I}=[\frac{1}{2},2]$
with $a^\mathrm{scale}$ and $b^\mathrm{scale}$ taking the value of the closest bound
if $a^\mathrm{scale}\notin\mathcal{I}$ or $b^\mathrm{scale}\notin\mathcal{I}$, respectively.

Finally, all parameters of the material model are updated for the upcoming
unloading step $k+1$ according to
\begin{equation} \label{eq:update}
  a^{k+1}_{(\bullet)}=a^\mathrm{scale}\cdot a^{k}_{(\bullet)},\quad
  b^{k+1}_{(\bullet)}=b^\mathrm{scale}\cdot b^{k}_{(\bullet)}.
\end{equation}
Note that all discussed material models require that $a^{k}_{(\bullet)}>0$ and $b^{k}_{(\bullet)}>0$ for all $k$ and hence $a^\mathrm{scale}$ and $b^\mathrm{scale}$ have to be positive,
which is guaranteed by design.
Additionally, we point out that for all simulations where the initial loading function was concave the presented algorithm still converged in a satisfactory manner.

To verify the proposed MFF approach, a more costly cost function based fitting (CFF) approach was established and applied to the available patient data. For details see~\ref{app:methods-cff}.
%------------------------------------------------------------------------------------------------
\subsubsection{Error estimates and goodness of fit}\label{sec:error_estimates}
To check for convergence, detect potential stagnation, and measure goodness of fit of the proposed method, several error estimates were introduced, which are described in detail below.
The difference of $V_0$ and  $V_\mathrm{ed}$ between the simulated and the empirical Klotz EDPVR was quantified by the difference of the initial volumes
\begin{equation}
\label{eq:err-initial}
    r^{V_0}=\left|V_0^\mathrm{klotz}-V_0^\mathrm{sim}\right|,
\end{equation}
where $V_0^\mathrm{sim}$ is the cavity volume of the reference configuration $\vec{X}^k$ at the current unloading step $k$,
and the difference of end-diastolic volumes
\begin{equation}
\label{eq:err-ed}
    r^{\mathrm{ed}}=\left|V_\mathrm{ed}^\mathrm{dat}-V_\mathrm{ed}^\mathrm{sim}\right|,
\end{equation}
where $V_\mathrm{ed}^\mathrm{dat}$ is the measured end-diastolic volume and
$V_\mathrm{ed}^\mathrm{sim}$ is the cavitary volume of the inflated configuration $\vec{x}^k$.
For convergence we chose an error tolerance of $\epsilon^\mathrm{vol}=0.5\%$ of $V_\mathrm{ed}^\mathrm{dat}$ such that
$r^{\mathrm{ed}}<\epsilon^\mathrm{vol}$ and $r^{V_0}<\epsilon^\mathrm{vol}$.

The difference between the simulated and measured end-diastolic geometry was calculated by consideration of the maximal nodal error
\begin{equation}
\label{eq:err-maxnodal}
    r^{\lVert\bm{x}\rVert_\infty} =\max\limits_{i \in [1,N_\mathrm{nodes}]}
    \left\lVert\bm{x}_i^k-\bm{x}_i^\mathrm{dat}\right\rVert_2,
\end{equation}
where $\lVert\bullet\rVert_2$ stands for the $\ell_2$-norm and $N_\mathrm{nodes}$ represents the total number of nodes in the geometry.
To obtain convergence,  $r^{\lVert\bm{x}\rVert_\infty}$ was required to be smaller than \SI{0.1}{\mm}.
Finally, we define the error in parameter update between unloading iterations as
\begin{equation}
\label{eq:err-param}
    r^{\text{param}}=\max(a^\mathrm{scale}, b^\mathrm{scale}),
\end{equation}
which is used to detect stagnation of the algorithm and required to be smaller than $0.001$ for convergence.
%   r^{\left|\Delta \bm{x}\right|} = \frac{1}{N}
%    \sum\limits_{i}^N\lVert\bm{x}_i^\mathrm{sim}-\bm{x}_i^\mathrm{dat}\rVert_2,
%\end{equation}
%
%\begin{equation}
%    r^{\left|\Delta \bm{x}\right|} = \sqrt{\frac{\int_\Omega(\bm{x}^\mathrm{sim}-\bm{x}^\mathrm{dat})
%    \cdot(\bm{x}^\mathrm{sim}-\bm{x}^\mathrm{dat}) \,\mathrm{d} \Omega}
%    {\int_\Omega\mathrm{d} \Omega}}
%\end{equation}

To define the goodness of fit for the simulation outcome in terms of the fitted curve and the Klotz curve,
the relative difference of initial volumes
\begin{equation}
\label{eq:err-initial-rel}
    r^{V_0,\mathrm{rel}}=\frac{\left|V_0^\mathrm{klotz}-V_0^\mathrm{sim}\right|}{V_\mathrm{ed}^\mathrm{dat}}\cdot 10^2,
\end{equation}
and the relative area difference
\begin{equation}
\label{eq:err-area-rel}
    r^{A_n,\mathrm{rel}}=\frac{r^{A_n}}{A_\mathrm{klotz}}\cdot 10^2
\end{equation}
of the normalized curve to the area under the Klotz curve $A_\mathrm{klotz}$ were calculated.
The absolute area difference $r^{A_n}$ between two curves in the 2D space is computed by the method introduced by~\citet{Jekel2019similarity},
which positions quadrilaterals $q$ between two curves and subsequently calculates the sum of their areas
\[
  r^{A_n} = \sum_q A_q,
\]
using Gauss' area formula
\[
  A_q=\frac{1}{2}\left|x_1y_2+x_2y_3+x_3y_4+x_4y_1-x_2y_1-x_3y_2-x_4y_3-x_1y_4\right|,
\]
where $(x_i,y_i)$ represent the vertices of a quadrilateral.

\subsection{Numerical Framework}
The unloading and parameter estimation scheme was implemented in the FE framework Cardiac
Arrhythmia Research Package (CARP)~\cite{Vigmond2008solvers}.
The simulations were constrained by applying spring-like boundary conditions at the rim of the clipped aorta and over the epicardial surface at the apex of the LV, respectively,
see~\cite{Marx2020,Strocchi2020simulating}.
For all the passive inflation simulations we rely on solver methods established previously~\cite{Augustin2016anatomically} which have been verified in a N-version benchmark study~\cite{Land2015verification}.
In brief, the linear systems were solved by a generalized minimal residual method (GMRES)
with a relative error reduction of $\epsilon=10^{-8}$. Efficient preconditioning was
based on the library \emph{PETSc} (\url{https://www.mcs.anl.gov/petsc/}).
For all cases we used 100 loading steps to ramp up the pressure from $p=0$ to $p=p_\mathrm{ed}$.
To speed up computations, we limited the number of Newton steps to two for the passive inflation
simulations during the unloading scheme. This did not alter simulation results and is justified as we
were only estimating the unloaded reference configuration and material parameter sets.
For the MFF with the Levenberg–Marquardt least-squares algorithm we used Python V. 3.8 and packages NumPy V. 1.16.6 and SciPy V. 1.2.3.

To validate our results, we performed a passive inflation experiment that starts with the found stress-free reference configuration and the determined parameter set.
For this passive inflation we used a fully converging Newton algorithm with a relative $\ell_2$-norm error reduction of the residual of $\epsilon=10^{-6}$ and 100 loading steps.
All goodness of fit measurements in the sections below were calculated using the PV-curve of this
validation experiment.
\section{Results}\label{sec:results}
%
%\subsection{Fitting with the model function}
%The proposed MFF approach was applied to the patient cohort of $N=19$ cases, see \Cref{sec:patientdata}, and subsequently cases with the best and worst fit were identified, see~\Cref{sec:results_all_cases}.
%The methods performance using various constitutive laws, see \Cref{sec:diff-matlaws}, and FE types, see \Cref{sec:res-fe}, was investigated for these two example cases.
%
%------------------------------------------------------------------------------------------------------------------------------------------------------------
\subsection{LV patient cohort results}\label{sec:results_all_cases}
\begin{figure}[htbp]
\centering
\begin{subfigure}{.48\textwidth}
  \includegraphics[width=.95\linewidth]{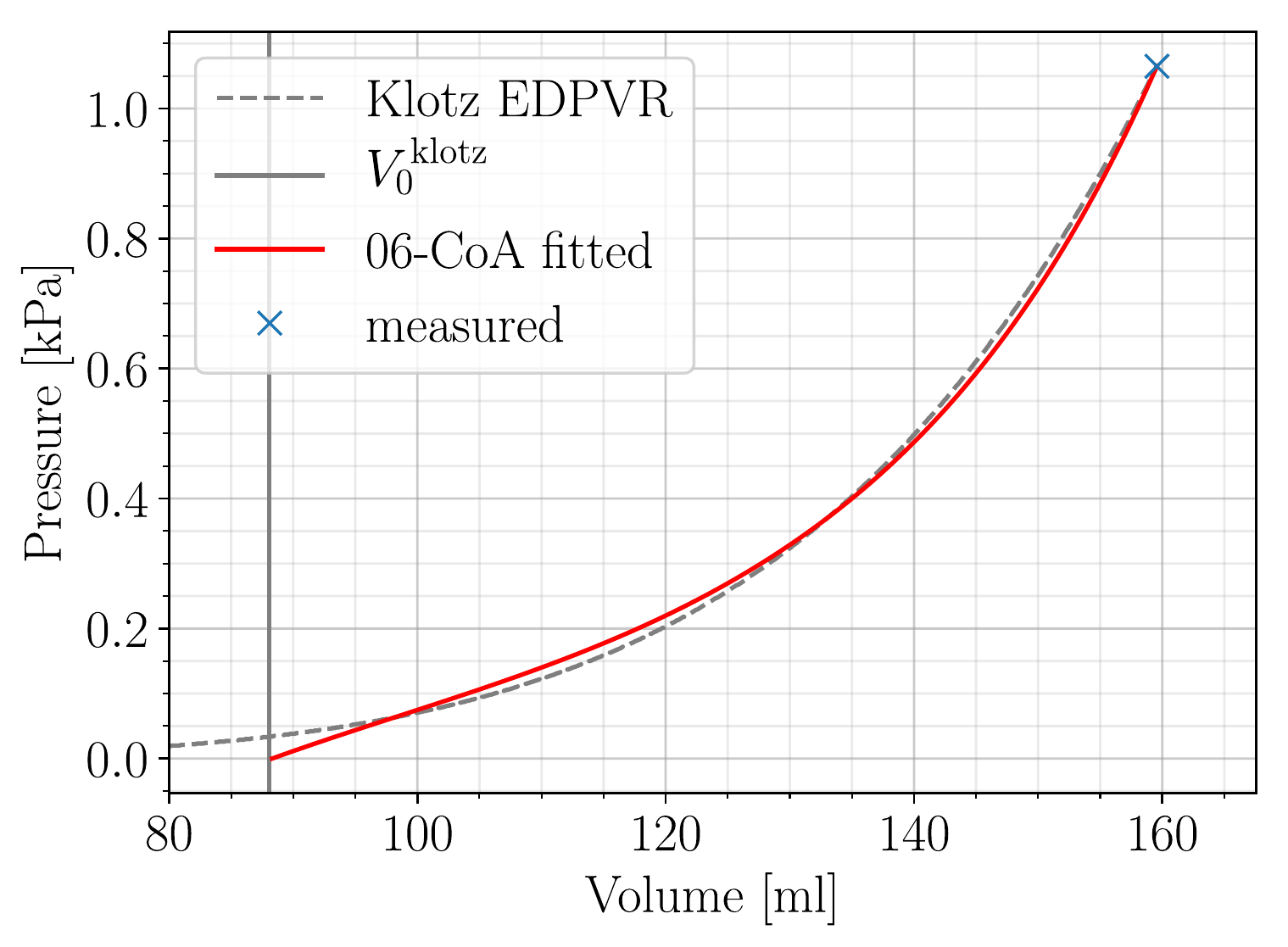}
  \label{fig:mffall_B62}
\end{subfigure}
\begin{subfigure}{.48\textwidth}
  \includegraphics[width=.95\linewidth]{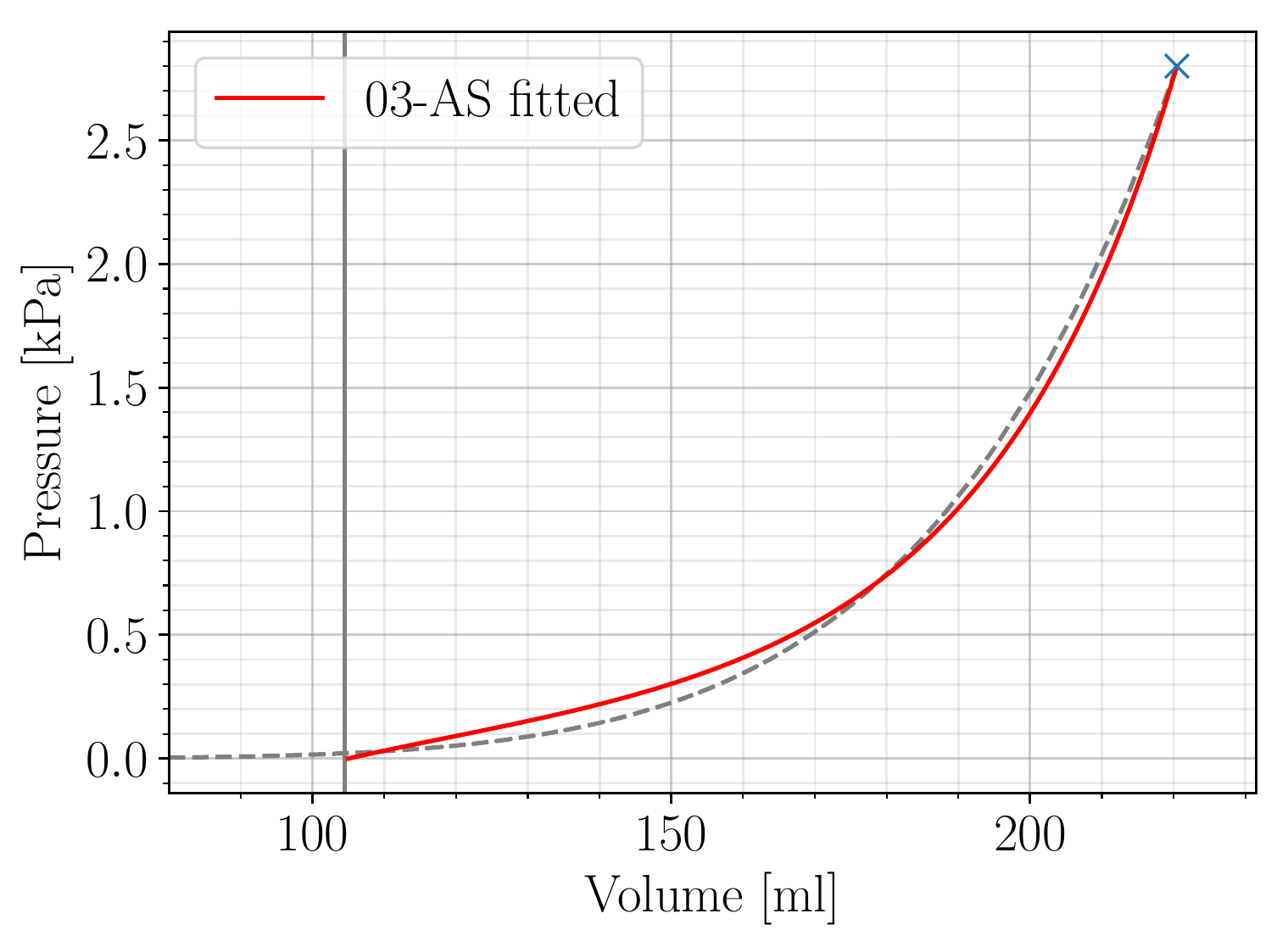}
  %\caption{B82-AS (worst fit)}
  \label{fig:mffall_B82}
\end{subfigure}
\caption{Fitting results of case 06-CoA (left) and 03-AS (right) are shown. The measured end-diastolic PV-point is marked in blue and used as a starting point for the Klotz EDPVR (dashed gray). $V_{0}^\mathrm{klotz}$ is shown as a vertical line in solid gray, while the fitted relation is visualized in solid red.}
\label{fig:mff-allcases}
\end{figure}

The MFF method as described in~\Cref{sec:model_function_fitting} was executed for all $N=19$ patient cases.
Parameters of the reduced HO constitutive law were fitted using initial values from the literature, see \Cref{tab:HolzapfelOgdenMaterialParameters}.
P1-P0-elements with the bulk modulus $\kappa=\SI{650}{\kPa}$ were utilized for all cases and
simulations were run on the Vienna Scientific Cluster 4 (VSC-4) using $96$ cores.
Given the convergence criteria as presented in~\Cref{sec:error_estimates}, the MFF algorithm converged for all $19$ cases and fitting results are summarized in~\Cref{tab:mff-allcases}.
Cases with best (06-CoA) and worst (03-AS) fit were distinguished by calculating the relative area difference $r^{A_n,\mathrm{rel}}$, see~\Cref{eq:err-area-rel}, and used for
further analysis in the sections below.
The Klotz curve and the simulated PV-curve for these two cases are visualized in~\Cref{fig:mff-allcases}.
Even for the worst case the fitted model
approximated the Klotz curve remarkably well,
with an almost exact match in terms of $V^\mathrm{klotz}_0$ and $V^\mathrm{dat}_\mathrm{ed}$.
Indeed, the MFF method achieved excellent goodness of fit for all cases.
\begin{table}[htbp]
\centering
\resizebox{0.75\linewidth}{!}{
    \begin{tabular}{lS[table-format=3.2]ccccS[table-format=3.2]}
\toprule
%            & \multicolumn{4}{c}{B62-CoA} & \multicolumn{4}{c}{B82-AS} \\
%\cmidrule(lr){2-5}\cmidrule(lr){6-9}
Case-ID   & \multicolumn{2}{c}{Input Parameters} & \multicolumn{2}{c}{Fitted Parameters}& \multicolumn{2}{c}{Goodness of Fit} \\
\cmidrule(lr){2-3}\cmidrule(lr){4-5}\cmidrule(lr){6-7}
&  $V_\mathrm{ed}^\mathrm{dat}$ & $p_\mathrm{ed}^\mathrm{dat}$ &  $a_{\mathrm{scale}}$ &  $b_{\mathrm{scale}}$  & $r^{V_0,\mathrm{rel}}$  & $r^{A_n,\mathrm{rel}}$ \\
& [\si{ml}] & [\si{kPa}]& & & [$\%V_\mathrm{ed}^\mathrm{dat}$] & [$\% A_\mathrm{klotz}$]  \\
\midrule
01-CoA & 155.65 & $1.75$ & $0.3418$ & $0.5988$ & $0.16$ & 10.00\\
02-CoA & 230.23 & $2.56$ & $0.5776$ & $0.5671$ & $0.18$ & 11.87\\
03-CoA & 96.48  & $0.72$ & $0.2111$ & $0.5054$ & $0.16$ & 8.34\\
04-CoA & 92.86  & $0.61$ & $0.1308$ & $0.6887$ & $0.16$ & 9.86\\
05-CoA & 152.83 & $1.49$ & $0.3278$ & $0.5458$ & $0.17$ & 9.69\\
06-CoA & 159.27 & $1.07$ & $0.2576$ & $0.6258$ & $0.13$ & 7.38\\
07-CoA & 123.72 & $1.09$ & $0.3651$ & $0.5387$ & $0.11$ & 8.39\\
\cmidrule(lr){2-7}
&  & &  & Mean (SD) & $0.15\,(0.03)$ & \centering 9.36\,(1.47)\tabularnewline
\midrule
01-AS  & 239.95 & $2.80$ & $0.3422$ & $0.4496$ & $0.15$ & 14.22\\
02-AS  & 275.31 & $2.80$ & $0.2953$ & $0.5647$ & $0.16$ & 14.29\\
03-AS  & 220.46 & $2.80$ & $0.2972$ & $0.5081$ & $0.20$ & 14.77\\
04-AS  & 97.27  & $2.80$ & $0.1550$ & $0.6813$ & $0.18$ & 14.71\\
05-AS  & 175.89 & $2.80$ & $0.2511$ & $0.6868$ & $0.17$ & 14.41\\
06-AS  & 237.83 & $2.80$ & $0.2947$ & $0.5526$ & $0.17$ & 14.55\\
07-AS  & 120.33 & $2.80$ & $0.2116$ & $0.5774$ & $0.18$ & 14.77\\
08-AS  & 281.70 & $2.80$ & $0.3204$ & $0.7124$ & $0.15$ & 14.10\\
09-AS  & 128.09 & $2.80$ & $0.2427$ & $0.5729$ & $0.16$ & 14.47\\
10-AS  & 208.75 & $2.80$ & $0.3303$ & $0.5043$ & $0.16$ & 14.41\\
11-AS  & 154.64 & $2.80$ & $0.2406$ & $0.6297$ & $0.16$ & 14.46\\
12-AS  & 169.14 & $2.80$ & $0.1840$ & $0.6567$ & $0.17$ & 14.54\\
\cmidrule(lr){2-7}
&  & &  & Mean (SD)& $0.17\,(0.02)$ & \centering 14.48\,(0.21)\tabularnewline
\bottomrule
\end{tabular}%
}
\caption{Fitting results for all $N=19$ cases are listed in terms of fitted scaling parameters, $a_\mathrm{scale}$ and  $b_\mathrm{scale}$, and measures of goodness of fit, $r^{V_0,\mathrm{rel}}$ and $r^{A_n,\mathrm{rel}}$, along with input parameters, $V_\mathrm{ed}^\mathrm{dat}$ and $p_\mathrm{ed}^\mathrm{dat}$, used to compute the Klotz EDPVR. Mean values and standard deviation (SD) of goodness of fit measures were computed separately for each etiology.}
\label{tab:mff-allcases}%
\end{table}%

\begin{figure}[htbp]
\centering
\begin{subfigure}{.48\textwidth}
  \centering
  \includegraphics[width=.95\linewidth]{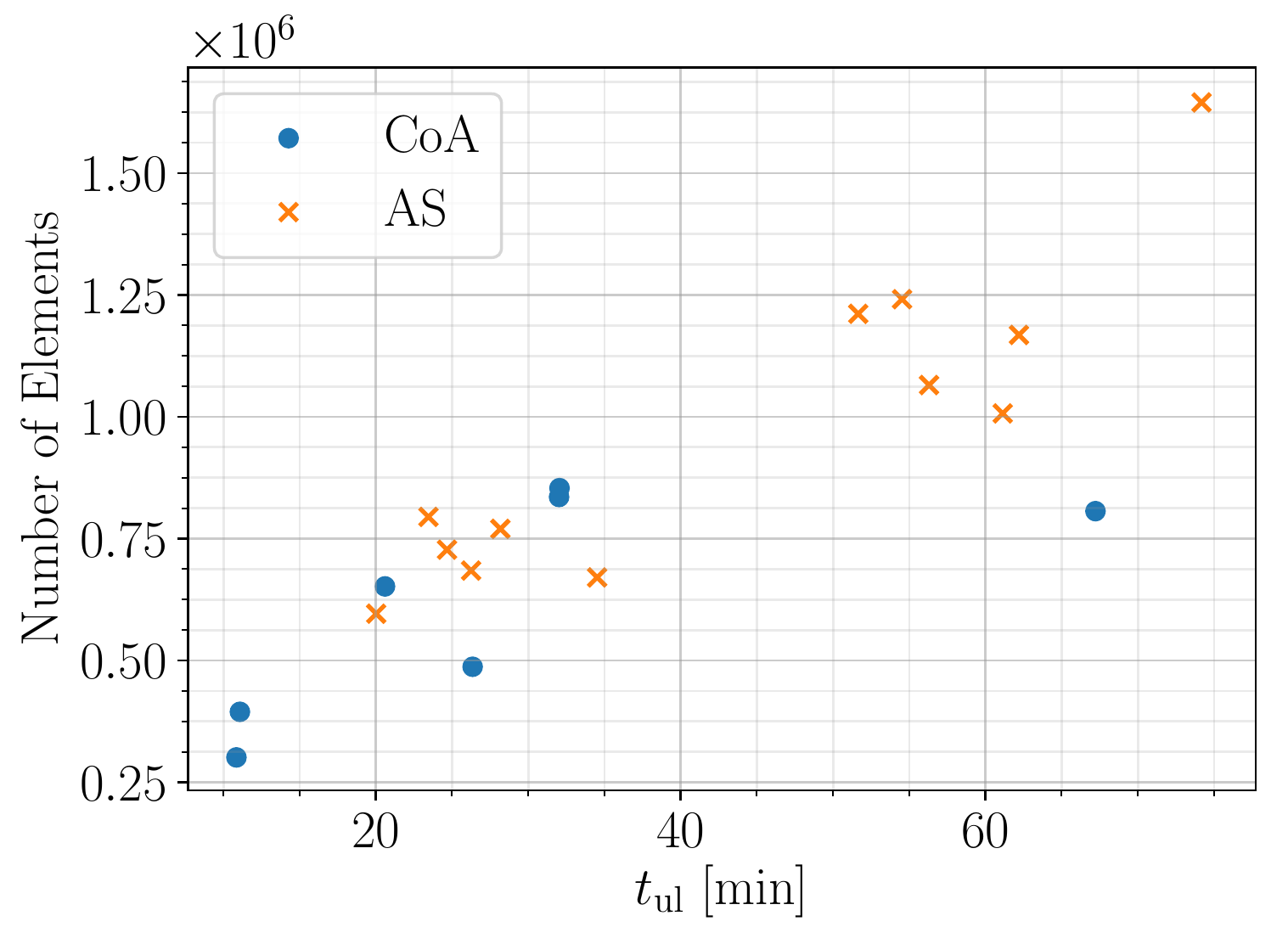}
\end{subfigure}
\begin{subfigure}{.48\textwidth}
  \centering
  \includegraphics[width=.95\linewidth]{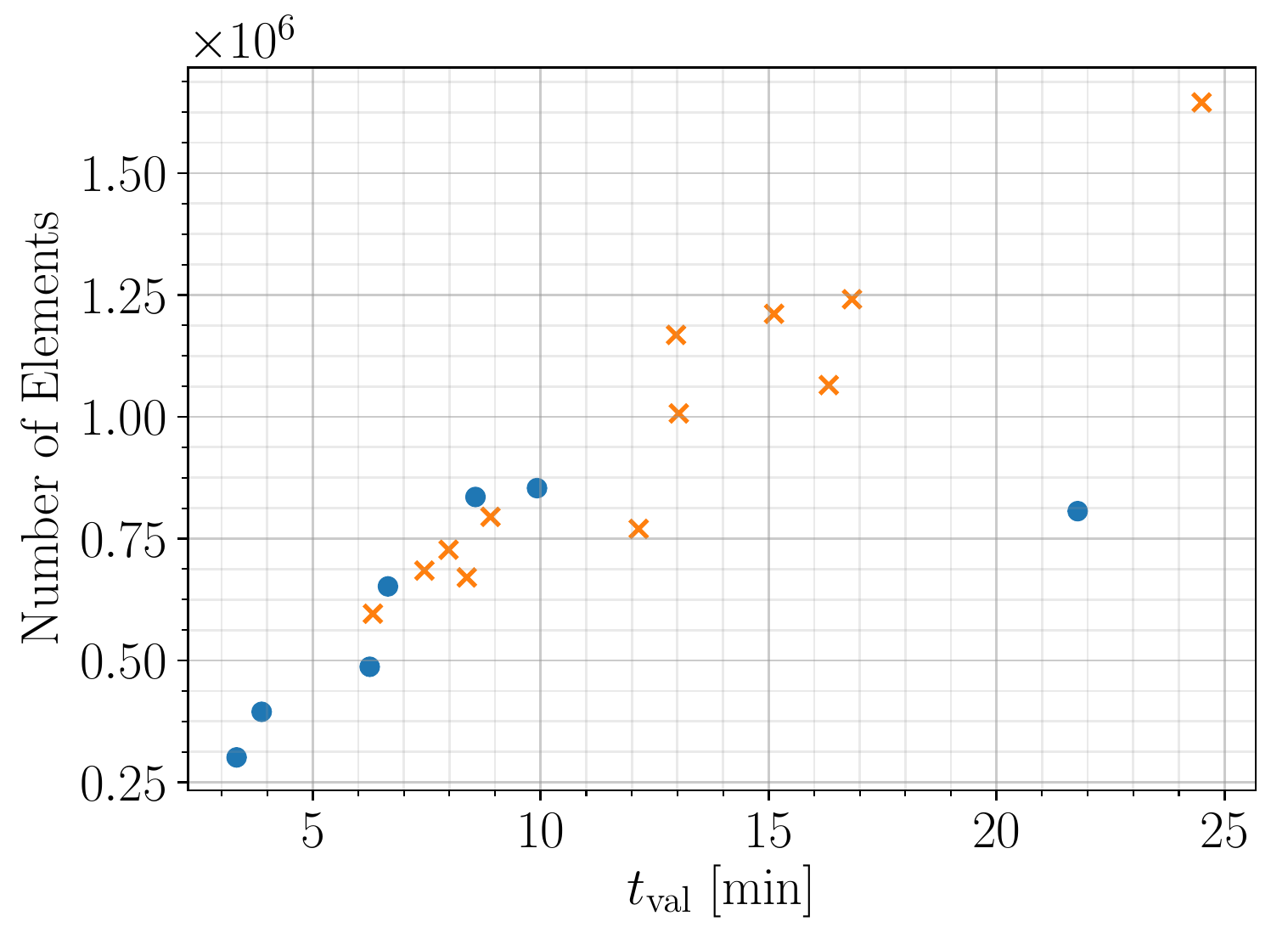}
\end{subfigure}
\caption{Relation of run-times for unloading $t_\mathrm{ul}$ (left) and validation $t_\mathrm{val}$ (right) to mesh size in terms of number of elements is visualized. Data from CoA Cases is shown as xxdjitsots, whereas data from AS Cases is marked as crosses.}
\label{fig:mff-allcases-meshsize}
\end{figure}

Run-times $t_\mathrm{ul}$ for the MFF method which includes unloading and parameter fitting were between \SI{10.85}{\min} and \SI{74.18}{\min}.
For the subsequent passive inflation experiment for validation the run-times $t_\mathrm{val}$ were between \SI{6.31}{\min} and \SI{24.50}{\min}.
As expected, computational cost increased with mesh size, see~\Cref{fig:mff-allcases-meshsize}.

\begin{figure}[htbp]
\centering
\begin{subfigure}{.48\textwidth}
  \includegraphics[width=0.95\linewidth]{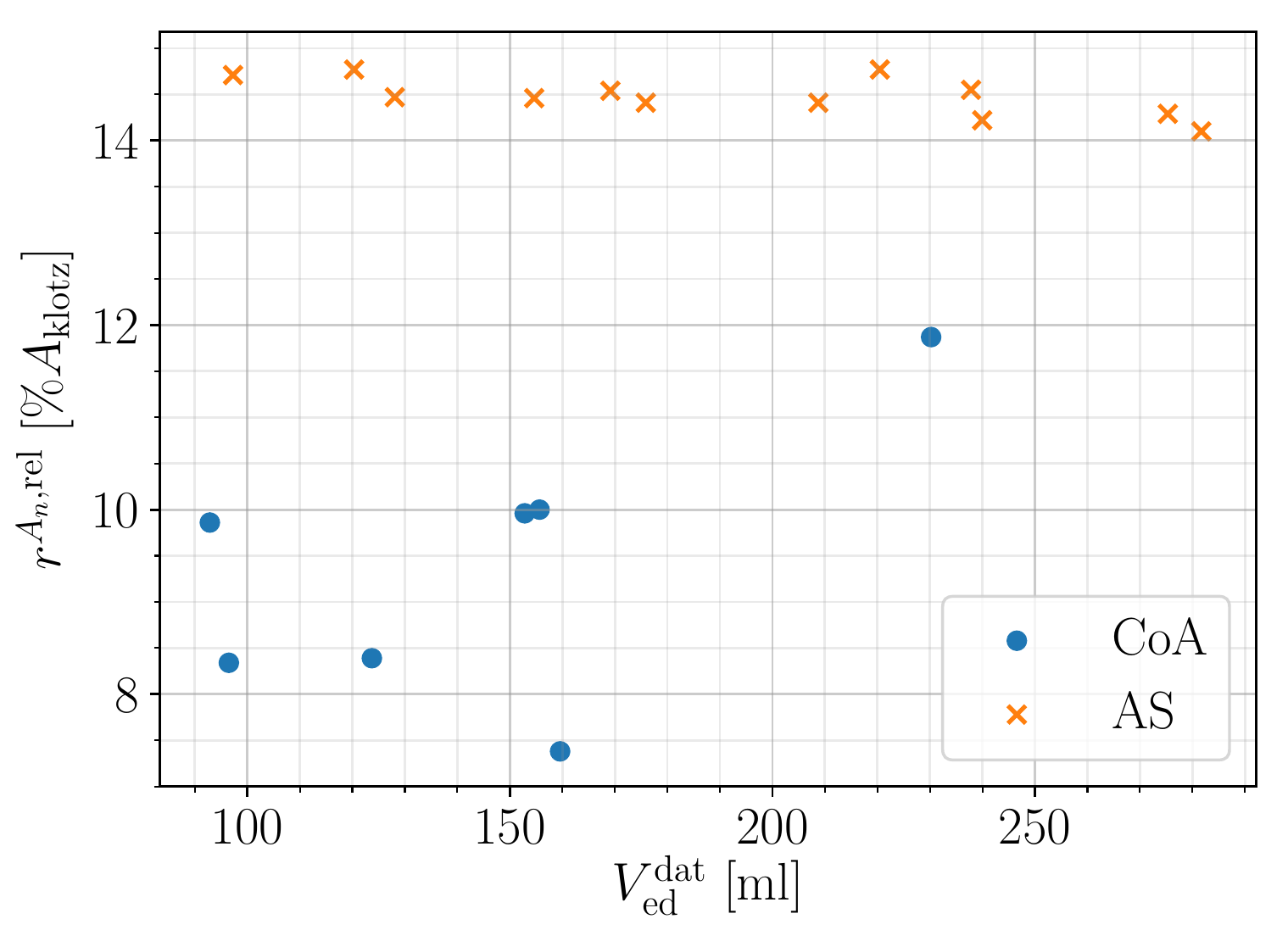}
\end{subfigure}
\begin{subfigure}{.48\textwidth}
  \includegraphics[width=0.95\linewidth]{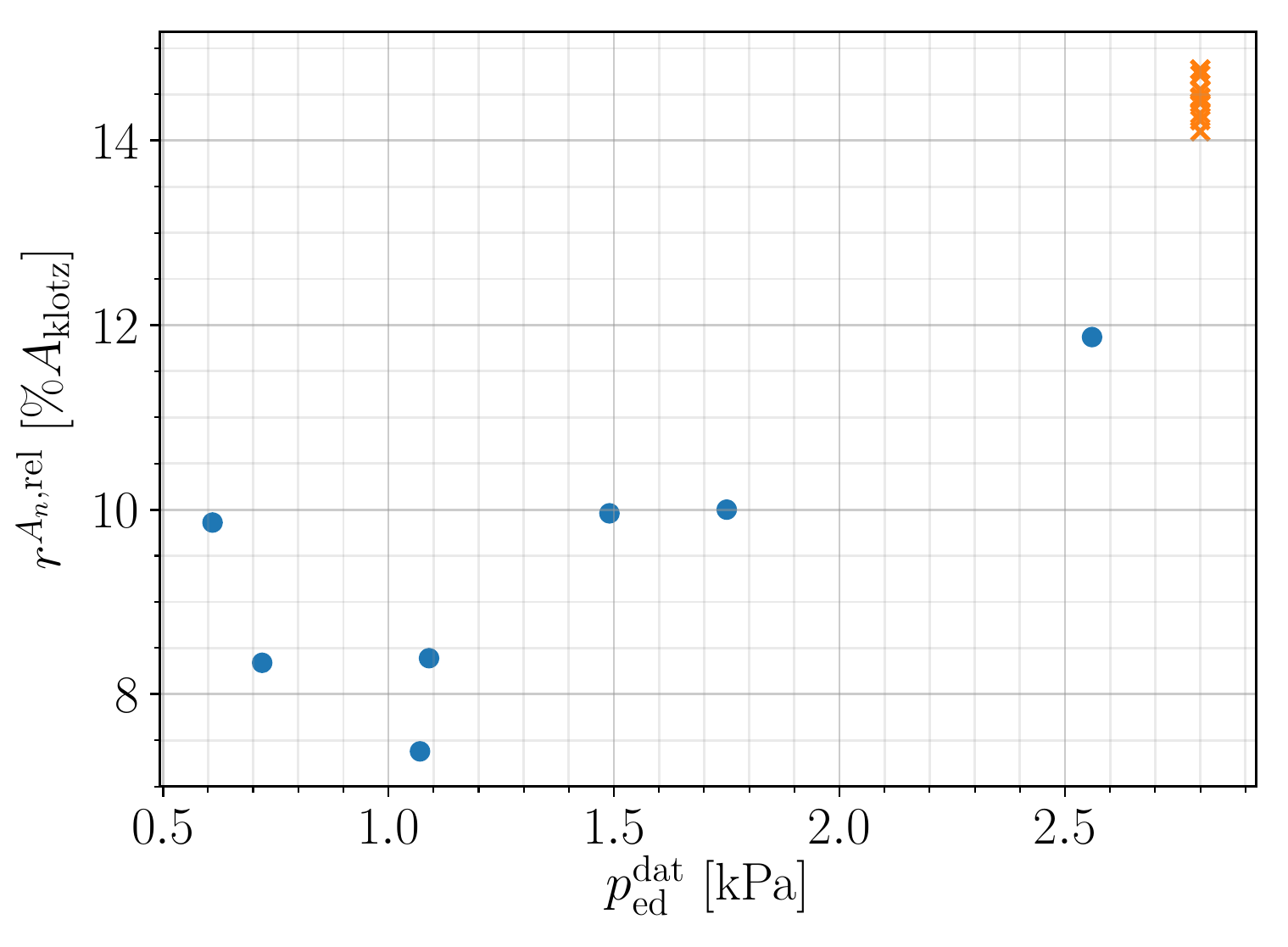}
\end{subfigure}
\caption{Influence of input parameters $V_\mathrm{ed}^\mathrm{dat}$ (left) and $p_\mathrm{ed}^\mathrm{dat}$ (right) on relative area difference
$r^{A_n,\mathrm{rel}}$ is shown. Data from CoA cases is marked as dots, whereas data from AS cases is marked as crosses.}
\label{fig:mff-allcases-input}
\end{figure}

Further, the influence of the input parameters $V_\mathrm{ed}^\mathrm{dat}$ and $p_\mathrm{ed}^\mathrm{dat}$ on goodness of fit,
more precisely $r^{A_n,\mathrm{rel}}$, is shown in~\Cref{fig:mff-allcases-input}. No trend can be recognized for different end-diastolic volumes $V_\mathrm{ed}^\mathrm{dat}$
while it appears that lower end-diastolic pressures $p_\mathrm{ed}^\mathrm{dat}$ might lead to better fits;
this is investigated further in \Cref{sec:sensitivity_analysis}.
%
% ----------------------------------------------------------------------------------------------------------
\FloatBarrier
\subsection{Results for different constitutive laws}
\label{sec:diff-matlaws}
The default constitutive law in~\Cref{sec:results_all_cases}, i.e., the reduced HO law with initial material parameters from~\citet{Guan2019AIC},
was chosen simply because this study is one of the most recent papers with fits to data from human ventricular myocardium.
In this section, we show that the method described above works equally well for a large variety of constitutive laws, in fact most of the laws that are currently used to model passive myocardium.

\begin{figure}[htbp]
\centering
%\begin{adjustbox}{minipage=\linewidth, scale=0.86}
    \begin{subfigure}{.45\textwidth}
        \centering
        % include first image
        \includegraphics[width=1.0\linewidth]{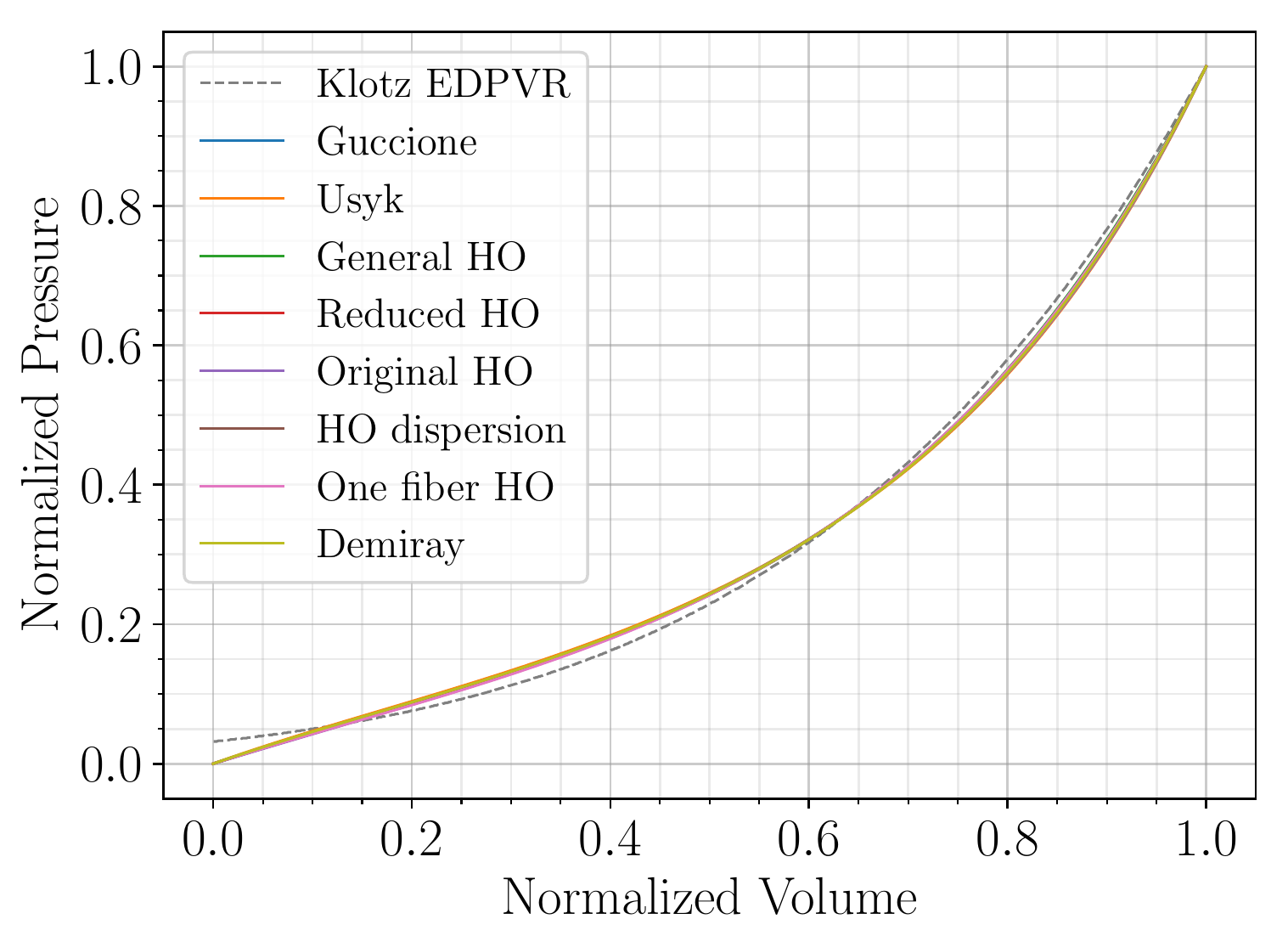}
        %\caption{Case 06-CoA (best fit)}
        %\label{fig:best}
    \end{subfigure}
    \begin{subfigure}{.45\textwidth}
        \centering
        % include second image
        \includegraphics[width=1.0\linewidth]{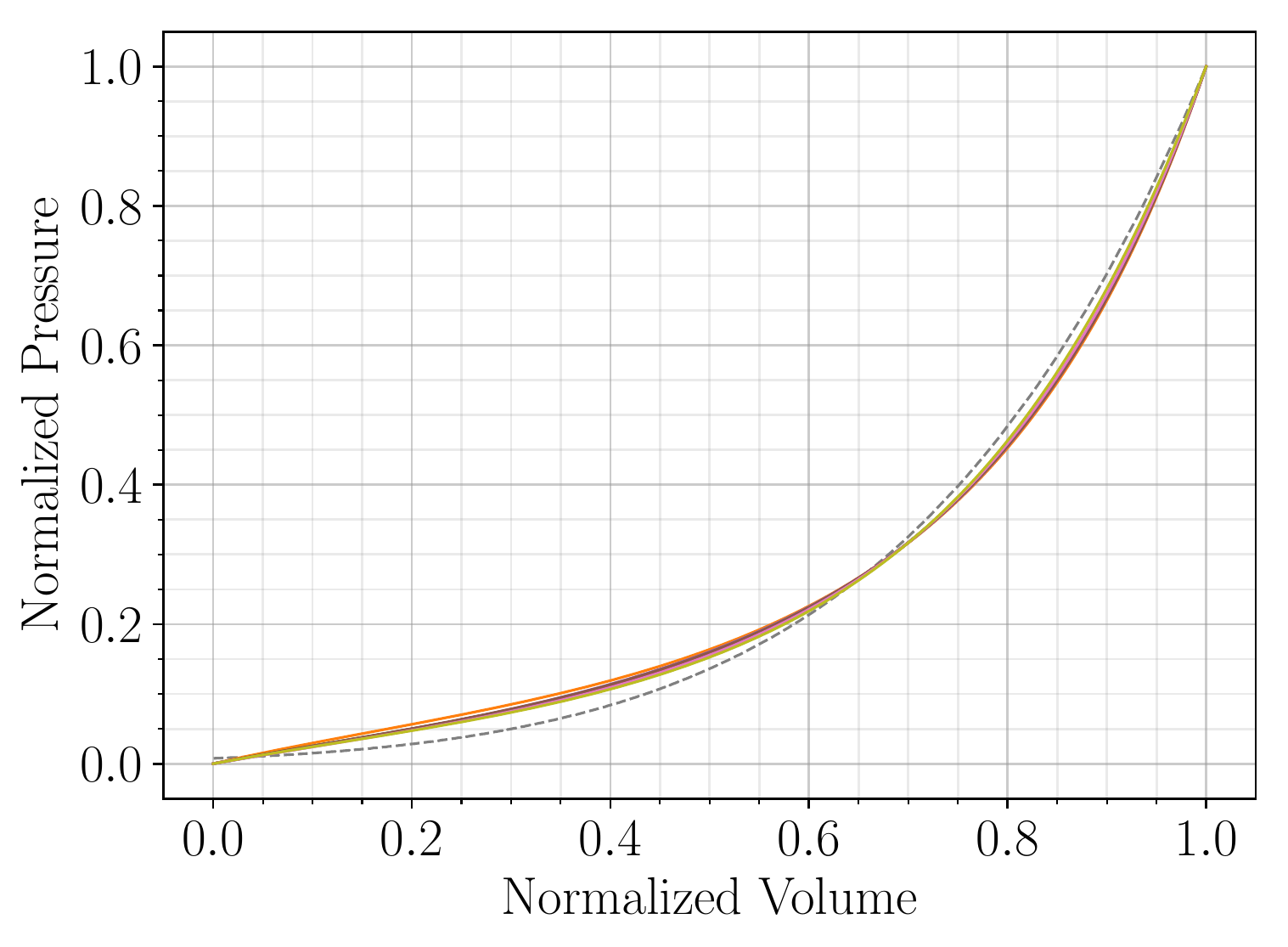}
        %\caption{Case 03-AS (worst fit)}
         %\label{fig:worst}
    \end{subfigure}
%\end{adjustbox}
\caption{Fitting results of case 06-CoA (left) and 03-AS (right) using different constitutive laws are shown. The normalized Klotz EDPVR (dashed gray) is visualized along with the respective normalized fitted curves.}
\label{fig:mff-material}
\end{figure}

To this end, we applied the MFF method using the different constitutive laws
listed in~\Cref{tab:singleFungMaterialParameters,tab:HolzapfelOgdenMaterialParameters}
for the case with best (06-CoA) and the worst (03-AS) fit in~\Cref{sec:results_all_cases}.
P1-P0-elements were used and all simulations were run on VSC4 using $96$ cores.
% with run-times for unloading including parameter fitting of $t_\mathrm{ul} = \SIrange{23.33}{151.63}{\min}$
%and run-times for the subsequent passive inflation experiments for validation of $t_\mathrm{val} = \SIrange{8.80}{20.33}{\min}$.
Normalized fitting results for cases 06-CoA and 03-AS are given in~\Cref{fig:mff-material} and~\Cref{tab:mff-material}
showing excellent fits and almost similar PV-curves for all material laws.
\begin{table}[htbp]
\resizebox{\linewidth}{!}{
    \begin{tabular}{lcccccccc}
\toprule
            & \multicolumn{4}{c}{ Case 06-CoA} & \multicolumn{4}{c}{Case 03-AS} \\
\cmidrule(lr){2-5}\cmidrule(lr){6-9}
   & \multicolumn{2}{c}{Fitted Parameters} & \multicolumn{2}{c}{Goodness of Fit}& \multicolumn{2}{c}{Fitted Parameters} & \multicolumn{2}{c}{Goodness of Fit} \\
\cmidrule(lr){2-3}\cmidrule(lr){4-5}\cmidrule(lr){6-7}\cmidrule(lr){8-9}
Material Model &  $a_{\mathrm{scale}}$ &  $b_{\mathrm{scale}}$  & $r^{V_0,\mathrm{rel}}$  & $r^{A_n,\mathrm{rel}}$  &  $a_{\mathrm{scale}}$ &  $b_{\mathrm{scale}}$  & $r^{V_0,\mathrm{rel}}$  & $r^{A_n,\mathrm{rel}}$ \\
  & & & [$\%V_\mathrm{ed}^\mathrm{dat}$] & [$\%A_\mathrm{klotz}$]& & & [$\%V_\mathrm{ed}^\mathrm{dat}$] & [$\%A_\mathrm{klotz}$] \\
\midrule
Guccione        & $0.0863$ & $1.8189$ & $0.16$ & $7.44$ & $0.1235$ & $1.4884$ & $0.21$ & $14.74$ \\
Usyk            & $0.0607$ & $3.9060$ & $0.19$ & $7.65$ & $0.1068$ & $2.7406$ & $0.24$ & $15.81$ \\
General HO      & $0.2630$ & $0.5892$ & $0.11$ & $7.27$ & $0.3034$ & $0.4702$ & $0.20$ & $14.76$ \\
Reduced HO      & $0.2576$ & $0.6258$ & $0.13$ & $7.38$ & $0.2972$ & $0.5081$ & $0.20$ & $14.77$ \\
Original HO     & $0.0473$ & $0.8441$ & $0.17$ & $7.59$ & $0.0615$ & $0.6707$ & $0.22$ & $14.79$ \\
HO dispersion   & $0.2973$ & $0.6781$ & $0.14$ & $7.65$ & $0.3554$ & $0.5354$ & $0.23$ & $14.83$ \\
One fiber HO    & $0.2649$ & $0.6382$ & $0.12$ & $7.74$ & $0.2934$ & $0.5518$ & $0.19$ & $14.50$ \\
Demiray         & $0.5050$ & $1.0502$ & $0.16$ & $7.49$ & $0.4955$ & $0.9411$ & $0.14$ & $14.52$ \\
\bottomrule
\end{tabular}%
}
\caption{Fitting results for case 06-CoA and 03-AS using different material laws are shown in terms of fitted scaling parameters, $a_{\mathrm{scale}}$ and $b_{\mathrm{scale}}$, and measures of goodness of fit, $r^{V_0,\mathrm{rel}}$ and $r^{A_n,\mathrm{rel}}$.}
\label{tab:mff-material}%
\end{table}%
Due to its simplicity and isotropic nature, the Demiray material model resulted in the lowest number of iterations and shortest run-times for both example cases.
More complex orthotropic models needed more iterations of the MFF algorithm and thus also run-times increased.
Overall, MFF simulations were tractable with a computational run-time below \num{2.5} hours for all experiments.
%modelsFor case 06-CoA, the use of the Usyk model resulted in longest run-times, more than twice as long as for the Demiray model,
%whereas for case 03-AS run-times for One fiber HO Model were six times as long as for the Demiray model.
% run-times:
% B62: t_ul = 23.33 -59.31 min, t_val = 8.80-16.73 min, fastest Demiray/slowest Usyk
% B82: t_ul = 25.45 -151.63 min, t_val = 11.18-20.33 min, fastest Demiray/ slowest HO1fiber
%
% ----------------------------------------------------------------------------------------------------------
\FloatBarrier
\subsubsection{Results for locking-free finite elements}
\label{sec:res-fe}
\begin{figure}[htbp]
\begin{subfigure}{.5\textwidth}
  \centering
  % include first image
  \includegraphics[width=.8\linewidth]{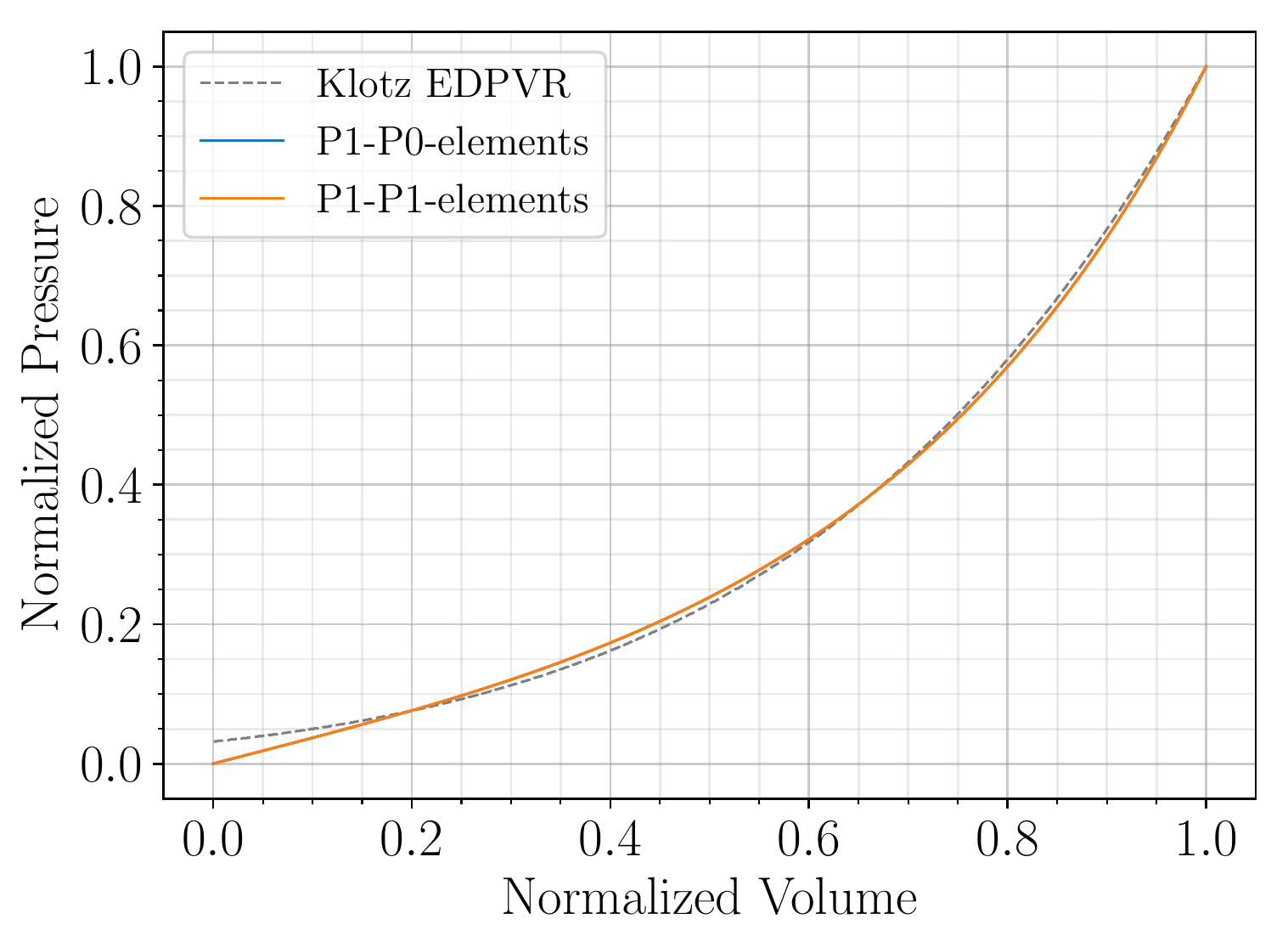}
  %\caption{Case 06-CoA normalized (best fit)}
  %\label{fig:mffall_B62_norm}
\end{subfigure}
\begin{subfigure}{.5\textwidth}
  \centering
  % include second image
  \includegraphics[width=.8\linewidth]{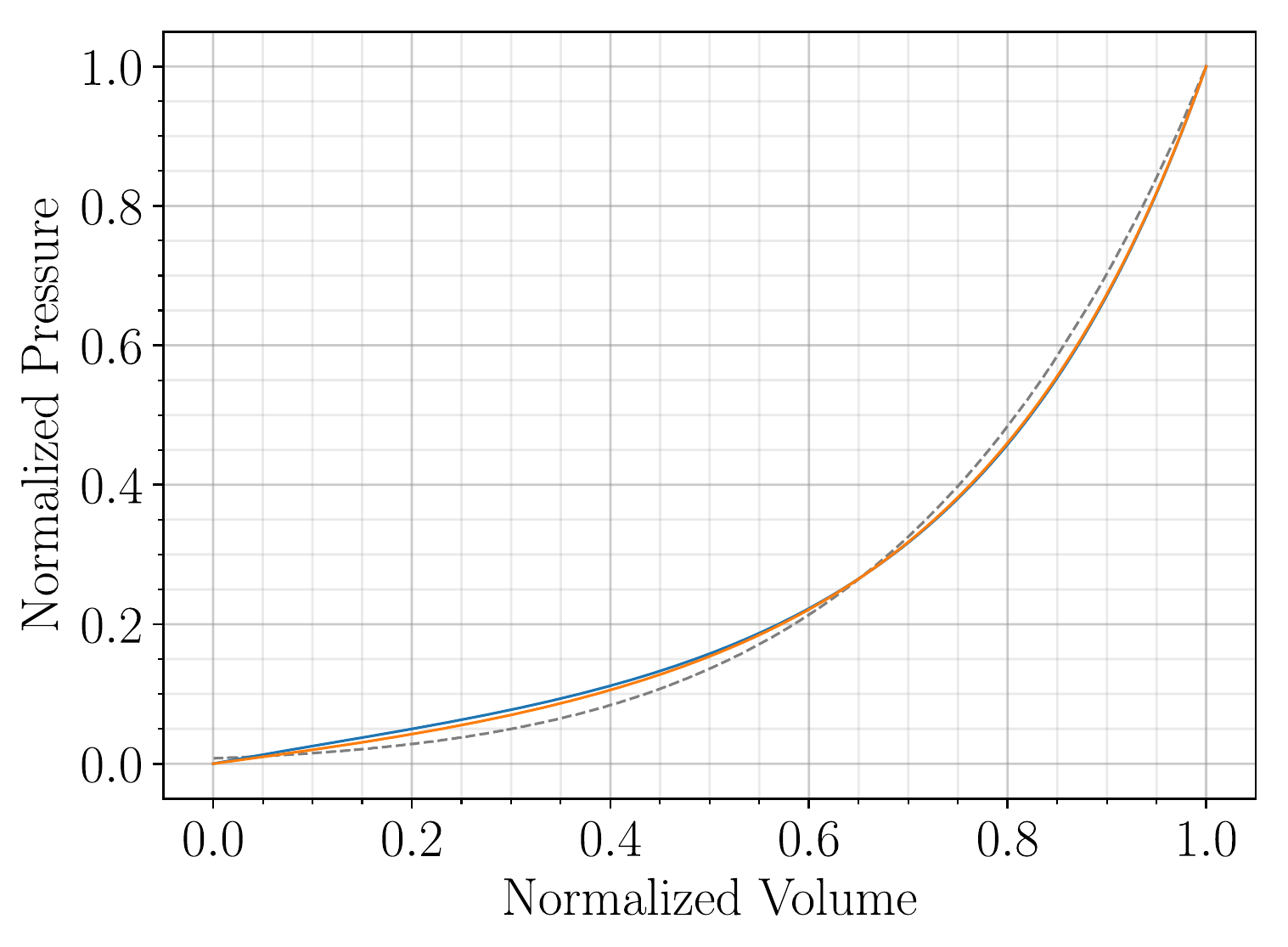}
  %\caption{Case 03-AS normalized (worst fit)}
  %\label{fig:mffall_B82_norm}
\end{subfigure}
\caption{Fitting results using P1-P0-elements (solid blue) and locking-free P1-P1-elements (solid orange) are compared for case 06-CoA (left) and 03-AS (right). The Klotz EDPVR is shown in dashed gray.}
\label{fig:mff-p1p1}
\end{figure}

It is well known that simple P1-P0-elements may suffer from locking effects and hence other FE formulations are required for certain applications where accurate stresses are essential. To show the capabilities of the MFF method in this scenario, we applied the algorithm to cases 06-CoA and 03-AS
using stabilized, locking-free P1-P1-elements~\cite{karabelas2020versatile} and an incompressible material, i.e., $1/\kappa=0$.
Normalized fitting results for the two example cases are presented in relation to results obtained above using P1-P0-elements in~\Cref{fig:mff-p1p1} and~\Cref{tab:mff-p1p1}.
The PV-curves and the measures of goodness of fit $r^{V_0,\mathrm{rel}}$ and $r^{A_n,\mathrm{rel}}$ show that the method works equally well for stabilized P1-P1-elements.
Looking at the parameter values in~\Cref{tab:mff-p1p1} we can see that $a_{\mathrm{scale}}$ is smaller and $b_{\mathrm{scale}}$ is larger for P1-P0-elements compared to the locking-free elements.
This is anticipated as the fitting compensates for a certain degree of locking in the P1-P0-element formulation and thus it gives parameters that correspond to a softer material.
Unsurprisingly, run-times were significantly longer when a stabilized P1-P1 FE formulation was used:
\SI{32}{\min} vs \SI{246}{\min} for case 06-CoA and
\SI{61}{\min} vs \SI{364}{\min} for case 03-AS,
 whereas iteration numbers of the MFF algorithm were similar.

 We could achieve similar results with locking-free MINI elements as introduced
 by~\citet{karabelas2020versatile}; only run-times varied a bit compared to the stabilized P1-P1 formulation.
\begin{table}[htbp] \small
\resizebox{\linewidth}{!}{
    \begin{tabular}{lcccccccc}
\toprule
& \multicolumn{4}{c}{Case 06-CoA} & \multicolumn{4}{c}{Case 03-AS} \\
\cmidrule(lr){2-5}\cmidrule(lr){6-9}
 & \multicolumn{2}{c}{Fitted Parameters}& \multicolumn{2}{c}{Goodness of Fit} & \multicolumn{2}{c}{Fitted Parameters}& \multicolumn{2}{c}{Goodness of Fit}\\
\cmidrule(lr){2-3}\cmidrule(lr){4-5}\cmidrule(lr){6-7}\cmidrule(lr){8-9}
Elem. Type &  $a_{\mathrm{scale}}$ &  $b_{\mathrm{scale}}$  & $r^{V_0,\mathrm{rel}}$  & $r^{A_n,\mathrm{rel}}$ &  $a_{\mathrm{scale}}$ &  $b_{\mathrm{scale}}$  & $r^{V_0,\mathrm{rel}}$  & $r^{A_n,\mathrm{rel}}$  \\
  & & & [$\%V_\mathrm{ed}^\mathrm{dat}$] & [$\% A_\mathrm{klotz}$] & & & [$\%V_\mathrm{ed}^\mathrm{dat}$] & [$\% A_\mathrm{klotz}$]\\
\midrule
P1-P0  & $0.2576$ & $0.6258$ & $0.13$ & $7.38$ & $0.2972$ & $0.5081$ & $0.20$ & $14.77$\\
P1-P1 stab & $0.3800$ & $0.5386$ & $0.08$ & $8.36$ & $0.4260$ & $0.4512$ & $0.19$ & $13.93$\\
\bottomrule
\end{tabular}%
}
\caption{Fitting results for case 06-CoA and 03-AS using P1-P0-elements and
stabilized P1-P1-element are compared in terms of fitted scaling parameters and measures of goodness of fit.}
\label{tab:mff-p1p1}
\end{table}%
\FloatBarrier
%
% ----------------------------------------------------------------------------------------------------------
\subsection{Sensitivity Analysis}
\label{sec:sensitivity_analysis}
To show robustness of the proposed MFF methodology, a sensitivity analysis was performed to investigate the influence of clinical data uncertainty
in terms of the end-diastolic PV-point used for computation of the Klotz EDPVR, changes in myocardial fiber orientation, altered model parameters used as initial guess
and variations of the bulk modulus $\kappa$, respectively.
The sensitivity analysis was executed for example case 03-AS using the reduced HO material law and P1-P0-elements if not mentioned otherwise.
%
% ----------------------------------------------------------------------------------------------------------
\subsubsection{Influence of variations in the end-diastolic PV-point on fitting outcome}
\label{sec:sa-edp}
\begin{figure}[htbp]
\begin{subfigure}{.5\textwidth}
  \centering
  % include first image
  \includegraphics[width=.8\linewidth]{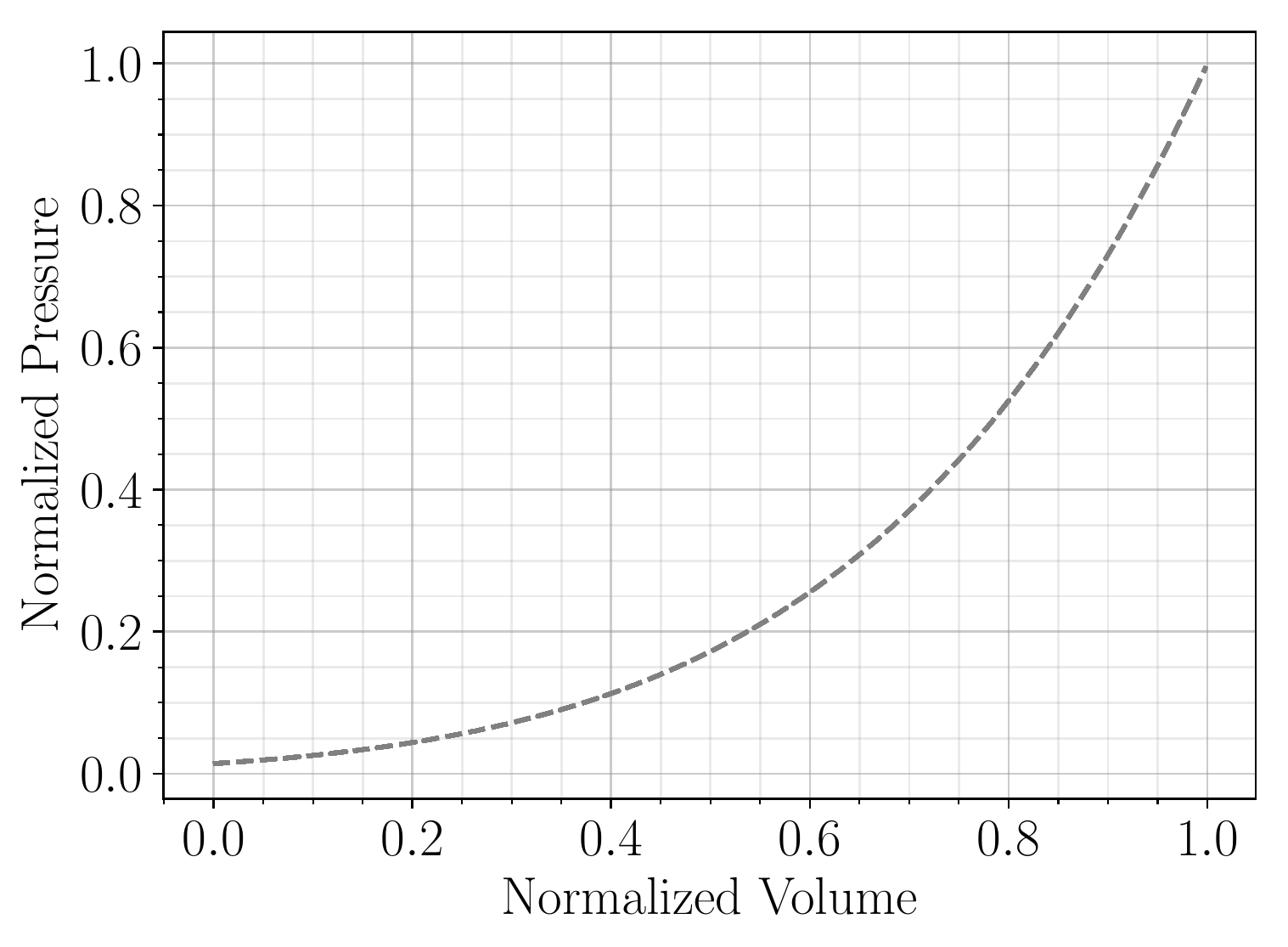}
  %\caption{Alternating $\mathrm{EDV_m}$ }
\end{subfigure}
\begin{subfigure}{.5\textwidth}
  \centering
  % include second image
  \includegraphics[width=.8\linewidth]{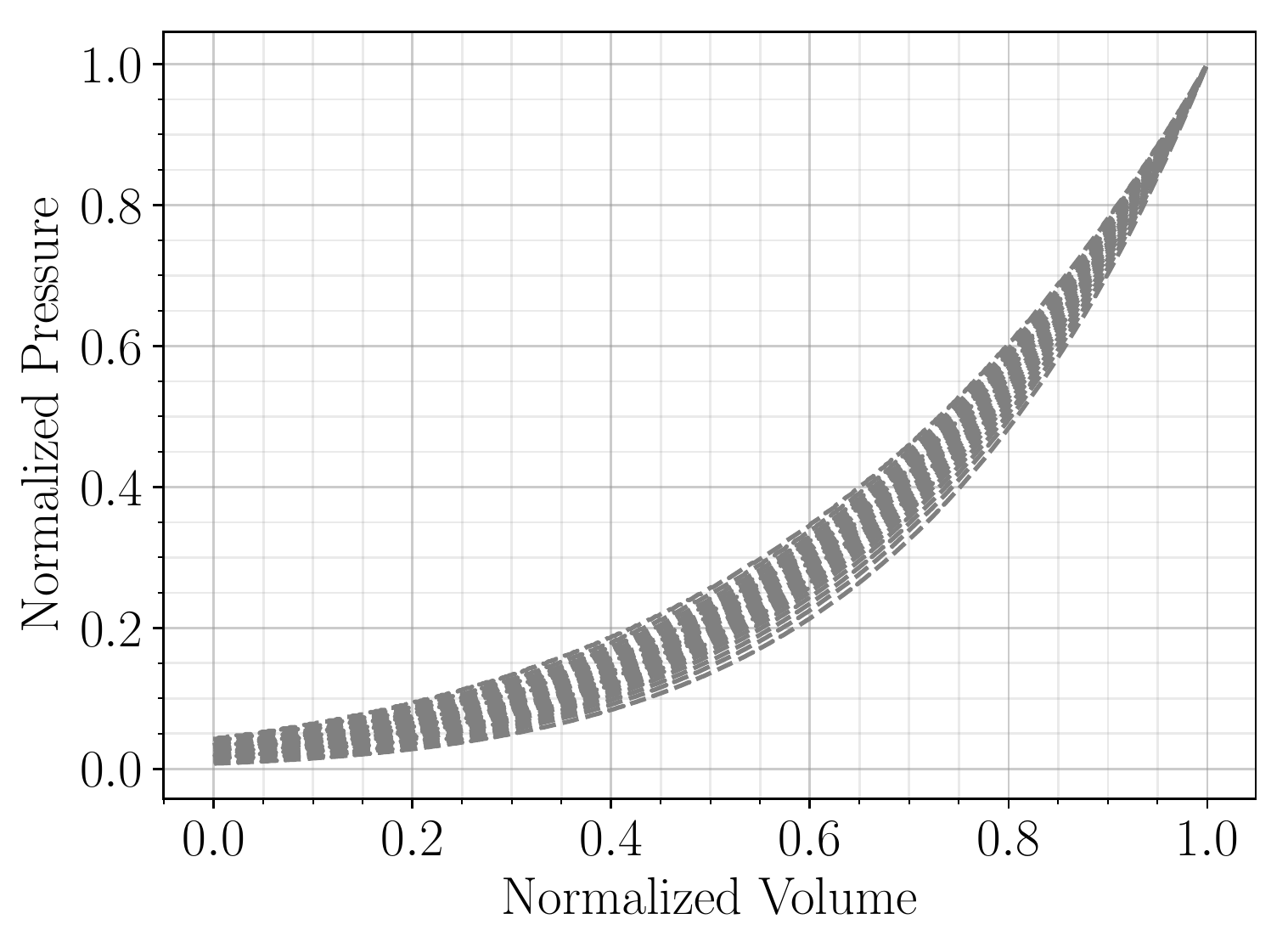}
  %\caption{Alternating $\mathrm{EDP_m}$}
\end{subfigure}
\caption{Behavior of the Klotz EDPVR relation is shown for varying end-diastolic PV-points. First, $V_\mathrm{ed}^\mathrm{dat}$ was varied while keeping $p_\mathrm{ed}^\mathrm{dat}$ constant and vice versa. Normalized curves match for alternating $V_\mathrm{ed}^\mathrm{dat}$ (left) while curves for alternating $p_\mathrm{ed}^\mathrm{dat}$ (right) differ in shape.}
\label{fig:sa-edp-klotz}
\end{figure}

The behavior of the Klotz curve was investigated changing the inputs $V_\mathrm{ed}^\mathrm{dat}$ and $p_\mathrm{ed}^\mathrm{dat}$ independently and subsequently comparing the normalized curves.
For that purpose, $n=20$ evenly spaced values in the range of \SIrange{93}{282}{\mL} for $V_\mathrm{ed}^\mathrm{dat}$ and \SIrange{0.61}{2.80}{\kPa} for $p_\mathrm{ed}^\mathrm{dat}$ were chosen;
range boundaries correspond to the respective minimum and maximum values in the patient cohort data.
For the variable held constant the mean values ($\overline{V}_\mathrm{ed}^\mathrm{dat}=\SI{175}{\mL}$ and $\overline{p}_\mathrm{ed}^\mathrm{dat}=\SI{2.26}{\kPa}$) obtained from the cohort data were taken.
First, $V_\mathrm{ed}^\mathrm{dat}$ was varied while keeping $p_\mathrm{ed}^\mathrm{dat}$ constant and vice versa. Normalized results of the generated curves are visualized in~\Cref{fig:sa-edp-klotz}.
It shows that curves generated with alternating $V_\mathrm{ed}^\mathrm{dat}$ match when normalized, whereas normalized curves resulting from alternating $p_\mathrm{ed}^\mathrm{dat}$ differ.

\begin{figure}[htbp]
\begin{subfigure}{.5\textwidth}
  \centering
  % include first image
  \includegraphics[width=.8\linewidth]{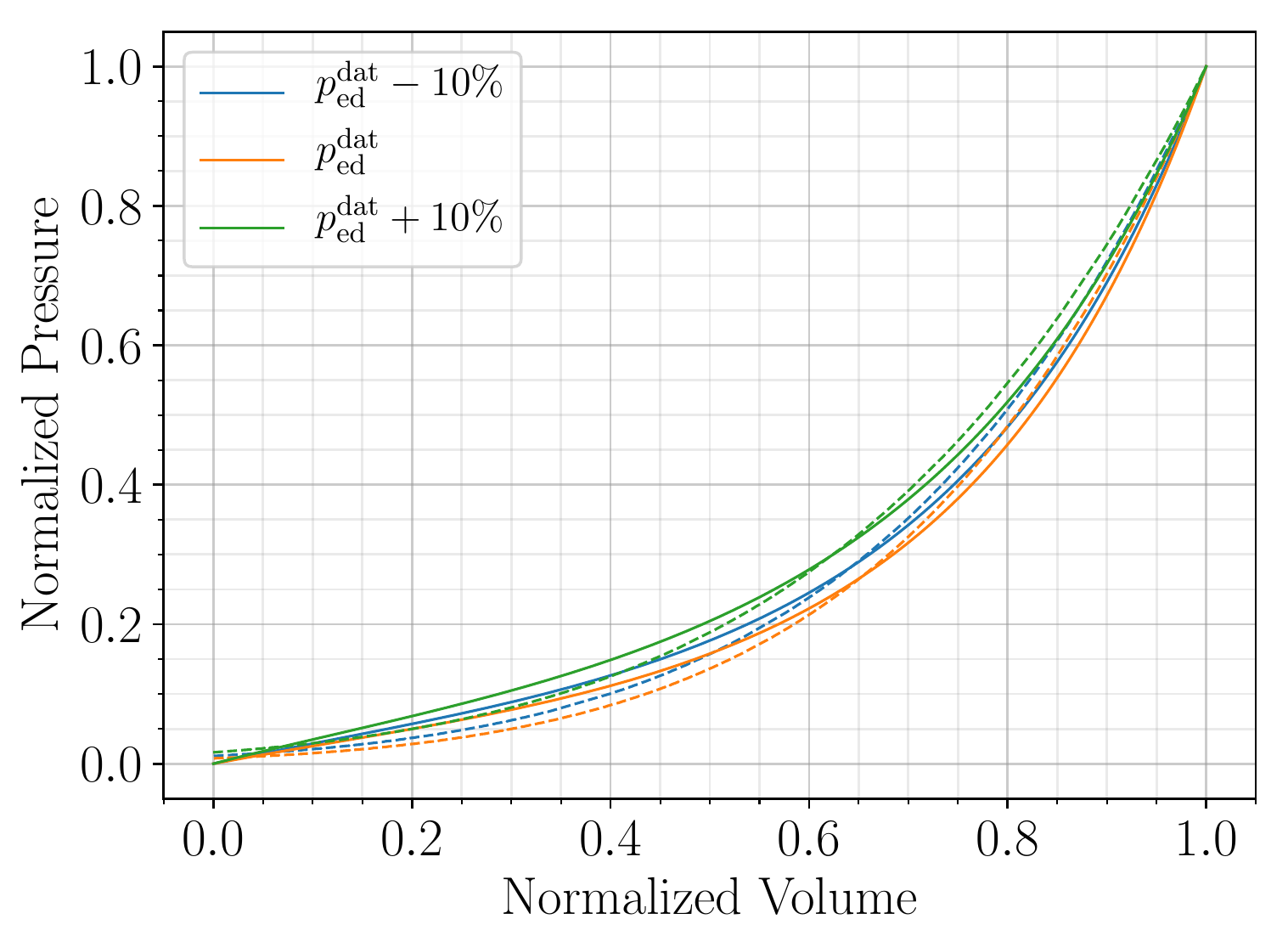}
  %\caption{Sensitivity to alternating $p_\mathrm{ed}$}
  %\label{fig:sa-edp}
\end{subfigure}
\begin{subfigure}{.5\textwidth}
  \centering
  % include second image
  \includegraphics[width=.8\linewidth]{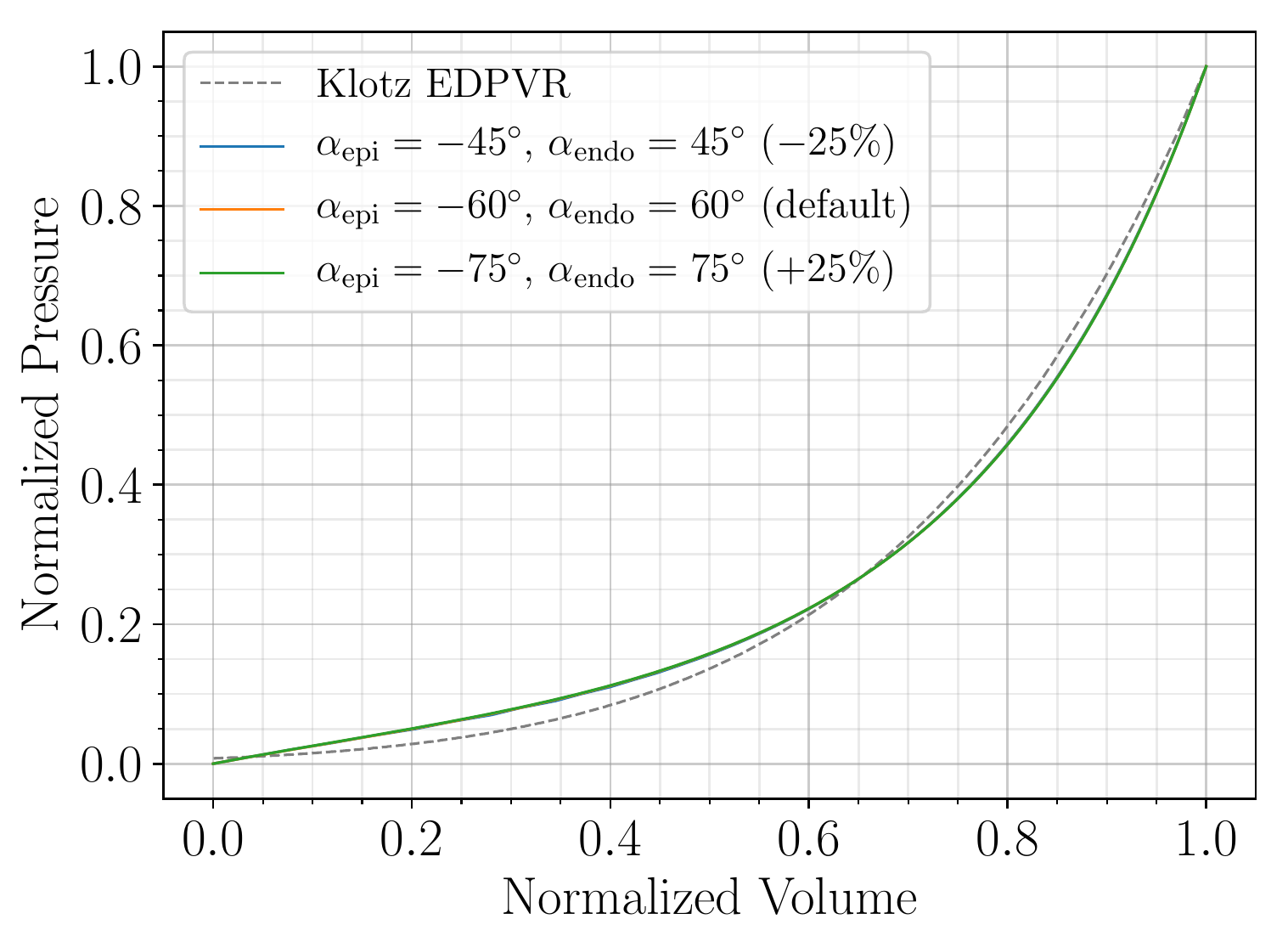}
  %\caption{Sensitivity to changing fiber distribution}
 %\label{fig:sa-fibers}
\end{subfigure}
\caption{Fitting results with respective Klotz EDPVR (dashed curves) for case 03-AS are compared for varying $p_\mathrm{ed}^\mathrm{dat}$ (left) and changing fiber angles (right).}
\label{fig:sa-fibers}
\end{figure}

For that reason, varying $V_\mathrm{ed}^\mathrm{dat}$ will have little effect on the goodness of fit of the MFF method. We could also observe this behavior previously in
\Cref{fig:mff-allcases-input}.
Hence, the sensitivity of the proposed MFF method was studied for varying $p_\mathrm{ed}^\mathrm{dat}$ only, considering deviations of \SI{\pm10}{\%}.
Results in~\Cref{fig:sa-fibers} and~\Cref{tab:sa-edp} suggest that fitting results are better for Klotz EDPVR that have higher normalized pressure values,
especially in the lower normalized volume region. Material parameters change for different values of $p_\mathrm{ed}^\mathrm{dat}$, but no
particular trend is recognizable.
\begin{table}[htbp] \small
 \centering
    \begin{tabular}{lccccc}
\toprule
 & & \multicolumn{2}{c}{Fitted Parameters}& \multicolumn{2}{c}{Goodness of Fit} \\
\cmidrule(lr){3-4}\cmidrule(lr){5-6}
$p_\mathrm{ed}^\mathrm{dat}$ & Deviation & $a_{\mathrm{scale}}$ &  $b_{\mathrm{scale}}$  & $r^{V_0,\mathrm{rel}}$  & $r^{A_n,\mathrm{rel}}$  \\
$[\si{\kilo\Pa}]$ & & & & [$\%V_\mathrm{ed}^\mathrm{dat}$] & [$\% A_\mathrm{klotz}$] \\
\midrule
$2.52$ & $-10\%$  & $0.3410$ & $0.4985$ & $0.21$ & $12.34$ \\
$2.80$ & $0\%$    & $0.2972$ & $0.5081$ & $0.20$ & $14.77$ \\
$3.08$ & $+10\%$  & $0.4903$ & $0.3803$ & $0.22$ & $7.68$ \\
\bottomrule
\end{tabular}%
\caption{Fitting results for case 03-AS are compared for varying $p_\mathrm{ed}^\mathrm{dat}$ in terms of fitted parameters and measures of goodness of fit.}
\label{tab:sa-edp}
\end{table}%
%
% ----------------------------------------------------------------------------------------------------------
\subsubsection{Sensitivity to fiber orientation}
Default myofiber orientations varying from $-60^{\circ}$ epicardial, $\alpha_\mathrm{epi}$, to $60^{\circ}$ endocardial, $\alpha_\mathrm{endo}$,
were chosen and perturbed by $25\,\%$. This resulted in fiber angles of $-75^{\circ}$ and $-45^{\circ}$ in the outer wall
and $75^{\circ}$ and $45^{\circ}$ in the inner wall, respectively. Sheet and sheet-normal directions were adapted accordingly to preserve the orthogonal system of local fiber coordinates.
Results of fitting performance are visualized in \Cref{fig:sa-fibers}
and values of fitted parameters and measures of goodness of fit are summarized in \Cref{tab:sa-fibers}.
As expected, parameter values are changing with the fiber orientation, as myocyte directions have a great impact on local material stiffness.
Further, we can observe that the goodness of fit of the MFF algorithm is almost indifferent to changes in
fiber orientation, leading to almost the same PV-curves in~\Cref{fig:sa-fibers}.
\begin{table}[htbp] \small
 \centering
    \begin{tabular}{lccccc}
\toprule
%& & \multicolumn{4}{c}{B82-AS} \\
%\cmidrule(lr){3-6}
 &  & \multicolumn{2}{c}{Fitted Parameters}& \multicolumn{2}{c}{Goodness of Fit} \\
\cmidrule(lr){3-4}\cmidrule(lr){5-6}
Fiber Orientation & Deviation &  $a_{\mathrm{scale}}$ &  $b_{\mathrm{scale}}$  & $r^{V_0,\mathrm{rel}}$  & $r^{A_n,\mathrm{rel}}$  \\
 & & & & [$\%V_\mathrm{ed}^\mathrm{dat}$] & [$\% A_\mathrm{klotz}$] \\
\midrule
$\alpha_\mathrm{epi}=-45^{\circ}$, $\alpha_\mathrm{endo}=45^{\circ}$ & $-25\%$ & $0.2768$ & $0.5005$ & $0.19$ & $14.68$ \\
$\alpha_\mathrm{epi}=-60^{\circ}$, $\alpha_\mathrm{endo}=60^{\circ}$  & $0\%$   & $0.2972$ & $0.5081$ & $0.20$ & $14.77$ \\
$\alpha_\mathrm{epi}=-75^{\circ}$, $\alpha_\mathrm{endo}=75^{\circ}$  & $+25\%$  & $0.3207$ & $0.5161$ & $0.20$ & $14.84$ \\
\bottomrule
\end{tabular}%
\caption{The influence of change in fiber orientation on fitting results in terms of fitted parameters and goodness of fit is listed.}
\label{tab:sa-fibers}
\end{table}%
%
% ----------------------------------------------------------------------------------------------------------
\subsubsection{Influence of initial model parameters on fitting outcome}
To show the robustness of the method, the influence of variations in initial model parameters of the reduced HO material model was investigated.
LHS from pyDOE V. 0.3.8 was used to create $n=10$ different sets of initial scaling parameters $\{a_\mathrm{scale},b_\mathrm{scale}\}$ in the interval $(0,1)$.
For all executed simulations the exact same parameter set and goodness of fit were obtained, also matching results with the default initial parameter set
$\{a_\mathrm{scale}=1,b_\mathrm{scale}=1\}$, see case 03-AS in \Cref{tab:mff-allcases}.
While this shows the high robustness of the MFF algorithm, we cannot prove uniqueness and the run-times of the simulations varied significantly: $t_\mathrm{ul} = \SIrange{39.48}{302.25}{\min}$;
depending on how close the initial parameters are to the final parameter set.
%
% ----------------------------------------------------------------------------------------------------------
\subsubsection{Influence of variations in bulk modulus on fitting outcome}
\begin{figure}[htbp]
\begin{subfigure}{.5\textwidth}
  \centering
  % include first image
  \includegraphics[width=.8\linewidth]{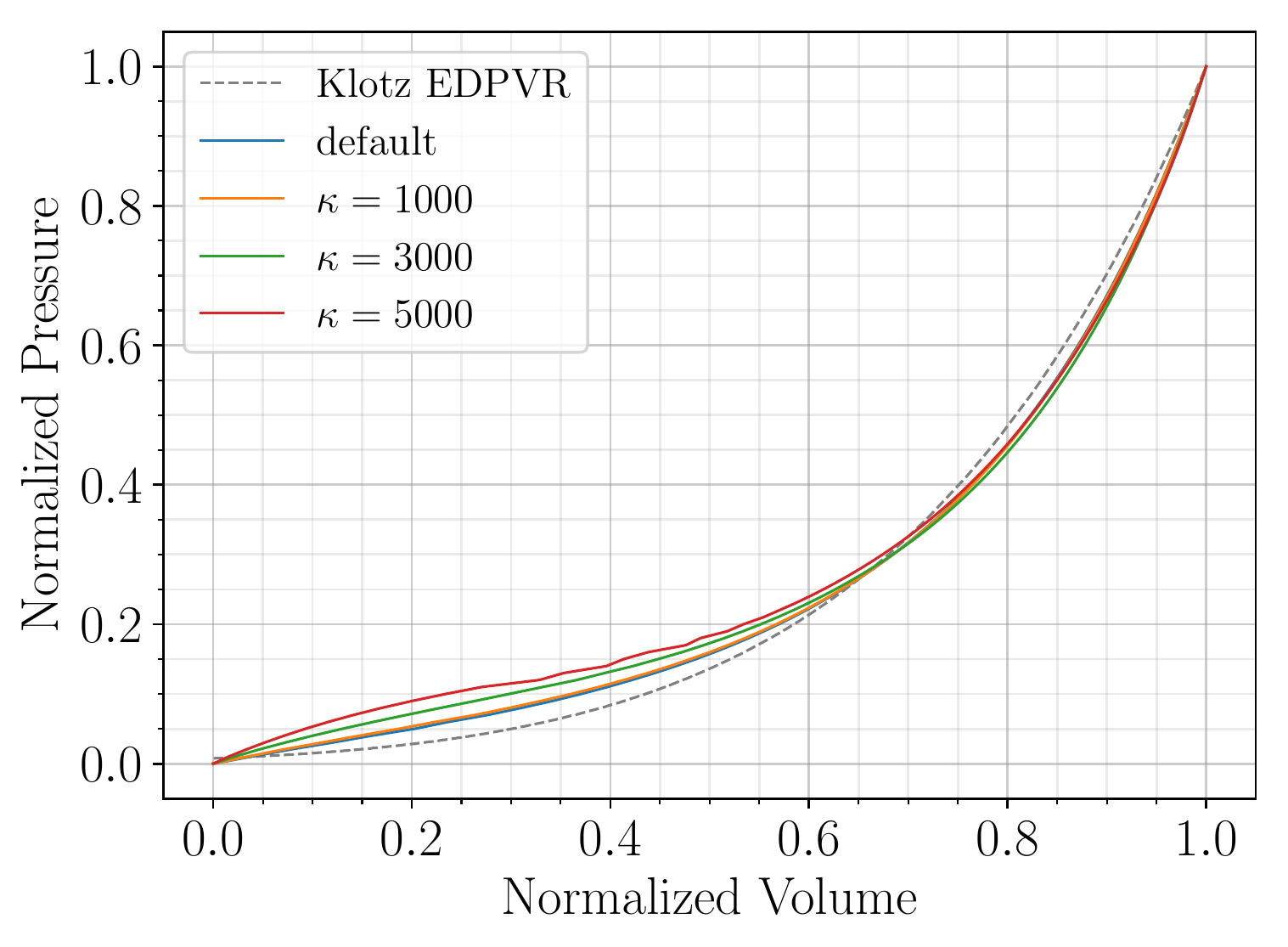}
  %\caption{P1-P0-elements}
  %\label{fig:mffall_B62_norm}
\end{subfigure}
\begin{subfigure}{.5\textwidth}
  \centering
  % include second image
  \includegraphics[width=.8\linewidth]{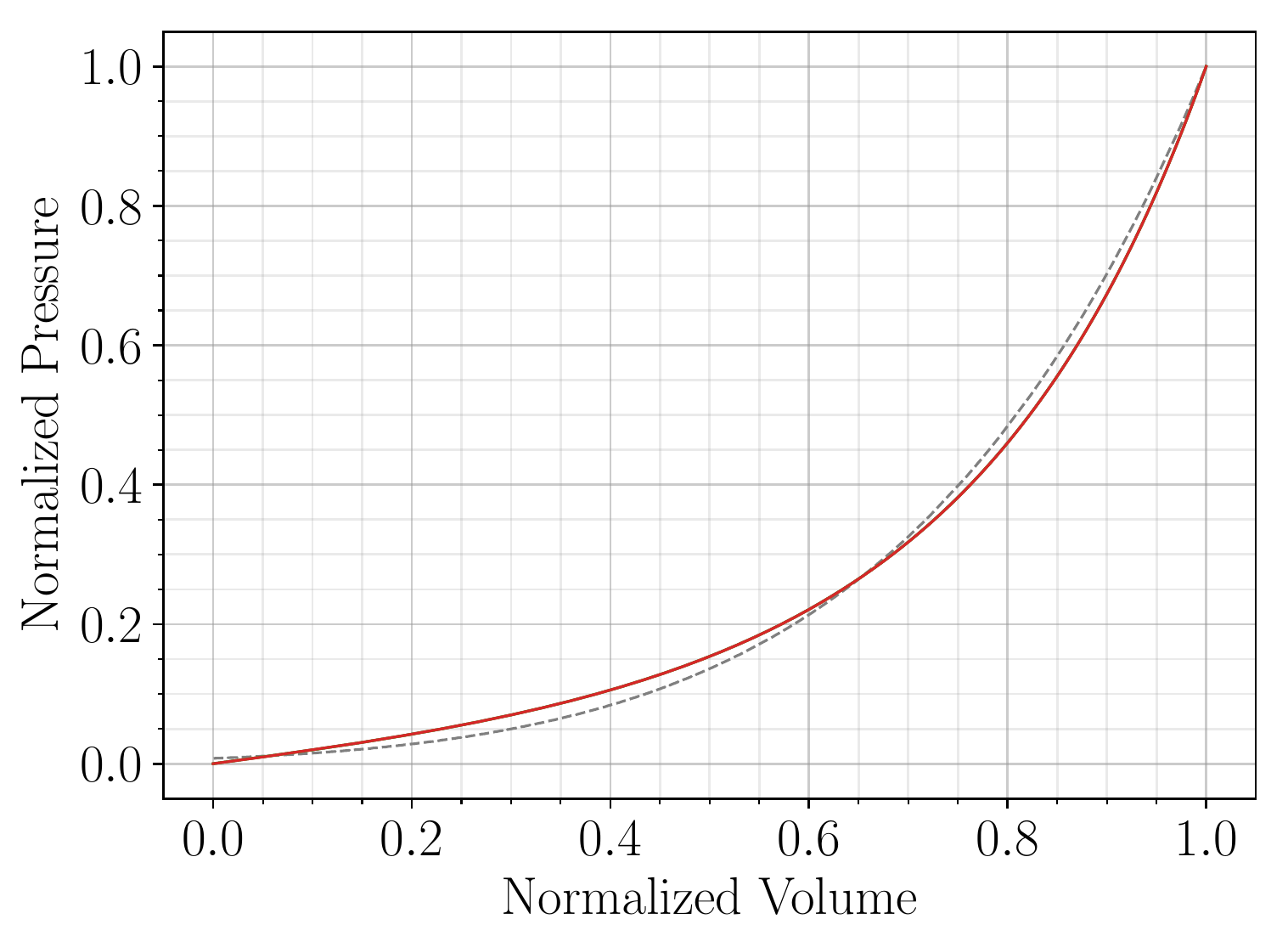}
  %\caption{P1-P1-elements}
  %\label{fig:mffall_B82_norm}
\end{subfigure}
\caption{Fitting results for case 03-AS are compared for different values of $\kappa$ using P1-P0-elements (left) and stabilized P1-P1-elements (right).}
\label{fig:sa-kappa}
\end{figure}

Finally, sensitivity of the method to changes of the bulk modulus $\kappa$ was studied using both P1-P0-elements and locking-free, stabilized P1-P1-elements.
Simulations were run for example case 03-AS for $\kappa=\{1000,3000,5000\}$\,\si{kPa}. Additionally, default values were chosen for P1-P0-elements ($\kappa=\SI{650}{kPa}$)
and for P1-P1-elements ($1/\kappa = 0$).
Results are visualized in \Cref{fig:sa-kappa} and summarized in \Cref{tab:sa-kappa}.
The forward simulations were not robust for P1-P0-elements and $\kappa=\SI{5000}{kPa}$ resulting in an irregular curve in~\Cref{fig:sa-kappa}(left).
This is a shortcoming of the penalty formulation and P1-P0-elements, not the MFF method itself. We can see that for P1-P0-elements the material parameters
vary greatly with the bulk modulus $\kappa$, compensating locking effects with material parameters that correspond to a softer material.
However, when using P1-P1-elements, the method was not sensitive to changes in $\kappa$.
Material parameters and goodness of fit are similar, which is a further indication of the robustness of the MFF algorithm.
% P1-P0: r_shape (ml): k=650:9.57, k=1000:10.86, k=3000:17.12, k=5000:22.46
% P1-P1: r_shape (ml) < 6.25 for all values of k
%
\begin{table}[htbp] \small
 \centering
    \begin{tabular}{lcccccccc}
\toprule
& \multicolumn{4}{c}{P1-P0-elements} & \multicolumn{4}{c}{P1-P1-elements}\\
\cmidrule(lr){2-5}\cmidrule(lr){6-9}
 & \multicolumn{2}{c}{Fitted Parameters}& \multicolumn{2}{c}{Goodness of Fit} & \multicolumn{2}{c}{Fitted Parameters}& \multicolumn{2}{c}{Goodness of Fit}\\
\cmidrule(lr){2-3}\cmidrule(lr){4-5}\cmidrule(lr){6-7}\cmidrule(lr){8-9}
$\kappa$ &  $a_{\mathrm{scale}}$ &  $b_{\mathrm{scale}}$  & $r^{V_0,\mathrm{rel}}$  & $r^{A_n,\mathrm{rel}}$ &  $a_{\mathrm{scale}}$ &  $b_{\mathrm{scale}}$  & $r^{V_0,\mathrm{rel}}$  & $r^{A_n,\mathrm{rel}}$ \\
 $[\si{\kilo\Pa}]$ & & & [$\%V_\mathrm{ed}^\mathrm{dat}$] & [$\% A_\mathrm{klotz}$] & & & [$\%V_\mathrm{ed}^\mathrm{dat}$] & [$\% A_\mathrm{klotz}$]\\
\midrule
default  & $0.2972$ & $0.5081$ & $0.20$ & $14.77$ & $0.4260$ & $0.4512$ & $0.19$ & $13.93$\\
$1000$ & $0.2614$ & $0.5199$ & $0.22$ & $15.26$ & $0.4266$ & $0.4511$ & $0.19$ & $13.93$ \\
$3000$ & $0.1149$ & $0.6263$ & $0.30$ & $18.08$ & $0.4264$ & $0.4511$ & $0.19$ & $13.92$ \\
$5000$ & $0.0340$ & $0.8113$ & $0.41$ & $20.15$ & $0.4266$ & $0.4511$ & $0.19$ & $13.94$ \\
\bottomrule
\end{tabular}%
\caption{Fitting results for case 03-AS are compared for different values of $\kappa$ using P1-P0-elements and stabilized P1-P1-elements.}
\label{tab:sa-kappa}
\end{table}%
%-----------------------------------------------------------------------------------------------------------
\FloatBarrier
\subsection{Comparison to fitting with cost functional and parameter space sampling}
\label{app:methods-cff}
%\textcolor{red}{Christoph: could be moved to appendix}\\
To compare and verify the results of the MFF approach we implemented an \emph{ad hoc} optimization
procedure, based on cost functional fitting (CFF), to find sets of material parameters.
To penalize deflections at lower and higher pressures and additionally penalize the difference to the stress-free
volume as determined by the Klotz relation~\eqref{eq:const-v0}, we chose the following cost functional:
\begin{align}
    J^{\mathrm{so2}}&=\frac{1}{2}\sum\limits_p  % {p=0}^{p_\mathrm{ed}}
    \left(\frac{V^\mathrm{klotz}(p)-V^\mathrm{sim}(p)}{V_\mathrm{ed}^\mathrm{dat}}\right)^2
    +\frac{1}{2}\sum\limits_V   %{V=V_0}^{V_\mathrm{ed}}
    \left(\frac{p^\mathrm{klotz}(V)-p^\mathrm{sim}(V)}{p_\mathrm{ed}^\mathrm{dat}}\right)^2 \nonumber\\
    &\quad+\gamma\left(\frac{V_0^\mathrm{klotz}-V_0^\mathrm{sim}}{V_\mathrm{ed}^\mathrm{dat}}\right)^2.
    \label{eq:cff}
\end{align}
Here, the first term is the sum of squared normalized volume differences, the second term is the sum of squared normalized pressure differences, and
the third term serves to further ensure that $V_0^\mathrm{klotz}$ and $V_0^\mathrm{sim}$ are close,
with $\gamma>0$ a weighting parameter; in all presented simulations we chose $\gamma=1$.
A downhill-Simplex based fitting (Nelder--Mead method) from SciPy was used to carry out the optimization, where we chose
an absolute error of $0.001$ in the parameter update between iterations to be acceptable for convergence.
We tried different varieties of CFF, the above described method with~\Cref{eq:cff} proved to be the most efficient.

%\paragraph{Sum of squared differences}
%\begin{align}
%    f^{\mathrm{so2}}=\sum\limits_p  % {p=0}^{p_\mathrm{ed}}
%    \left(\frac{V^\mathrm{klotz}(p)-V^\mathrm{sim}(p)}{V_\mathrm{ed}^\mathrm{dat}}\right)^2
%\end{align}
%\paragraph{Maximal deflection}
%\begin{equation}
%    f^{\max}=\frac{1}{2}\max\limits_p %{p\in[p_0,p_\mathrm{ed}]}
%      \left|\frac{V^\mathrm{klotz}(p)-V^\mathrm{sim}(p)}{V_\mathrm{ed}^\mathrm{dat}}\right|
%      + \frac{1}{2}\max\limits_V %{V\in[V_0,V_\mathrm{ed}]}
%      \left|\frac{p^\mathrm{klotz}(V)-p^\mathrm{sim}(V)}{V_\mathrm{ed}^\mathrm{dat}}\right|
%    + \left|\frac{V_0^\mathrm{klotz}(p)-V_0^\mathrm{sim}(p)}{V_\mathrm{ed}^\mathrm{dat}}\right|
%\end{equation}

\paragraph{Results} \label{app:results-cff}
The CFF approach was executed for the two example cases (06-CoA and 03-AS) with the reduced HO constitutive law,
P1-P0-elements, and the bulk modulus $\kappa=\SI{650}{\kPa}$.
We used default values for the material parameters as in~\Cref{tab:HolzapfelOgdenMaterialParameters}
scaled by a factor of 0.5 as we already expect from results in~\Cref{tab:mff-allcases} that material parameters from the
literature are in general too stiff; this scaling sped up computational times considerably.

Using the Nelder--Mead algorithm we optimized
(i) $n_\mathrm{opt}=1$ scaling parameter ($a_\mathrm{scale}$) which was applied to all material parameters $a_{(\bullet)}$, whereas all $b_{(\bullet)}$ parameters of the model were scaled with $0.5$;
(ii) $n_\mathrm{opt}=2$ scaling parameters ($a_\mathrm{scale}$, $b_\mathrm{scale}$), with $a_\mathrm{scale}$ applied to all $a_{(\bullet)}$ parameters and $b_\mathrm{scale}$ applied to all $b_{(\bullet)}$ parameters of the model; and
(iii) $n_\mathrm{opt}=3$ scaling parameters ($a_\mathrm{scale}$, $b^\mathrm{iso}_\mathrm{scale}$, $b^\mathrm{aniso}_\mathrm{scale}$), where $a_\mathrm{scale}$ was applied to all $a_{(\bullet)}$ parameters, while
$b$ in the isotropic contribution of the constitutive law was scaled with $b^\mathrm{iso}_\mathrm{scale}$ and all $b_{(\bullet)}$ parameters of the anisotropic contribution were scaled with $b^\mathrm{aniso}_\mathrm{scale}$.
We used bounds such that all scaling parameters are $>0.1$ to prevent that parts of the constitutive law are eliminated.
The case $n_\mathrm{opt}=4$, where additionally $a$ from the isotropic contribution and all $a_{(\bullet)}$ from the anisotropic contribution were scaled by two different parameters, was not converging for the patient-specific cases within
a reasonable amount of time (72 hours).
We performed the optimization using values from the literature, see \Cref{tab:HolzapfelOgdenMaterialParameters},
as starting values (no init) as well as using the MFF approach to generate initial guesses for the parameter scalings (init).

\begin{figure}[htbp]
\begin{subfigure}{.5\textwidth}
  \centering
  % include first image
  \includegraphics[width=.8\linewidth]{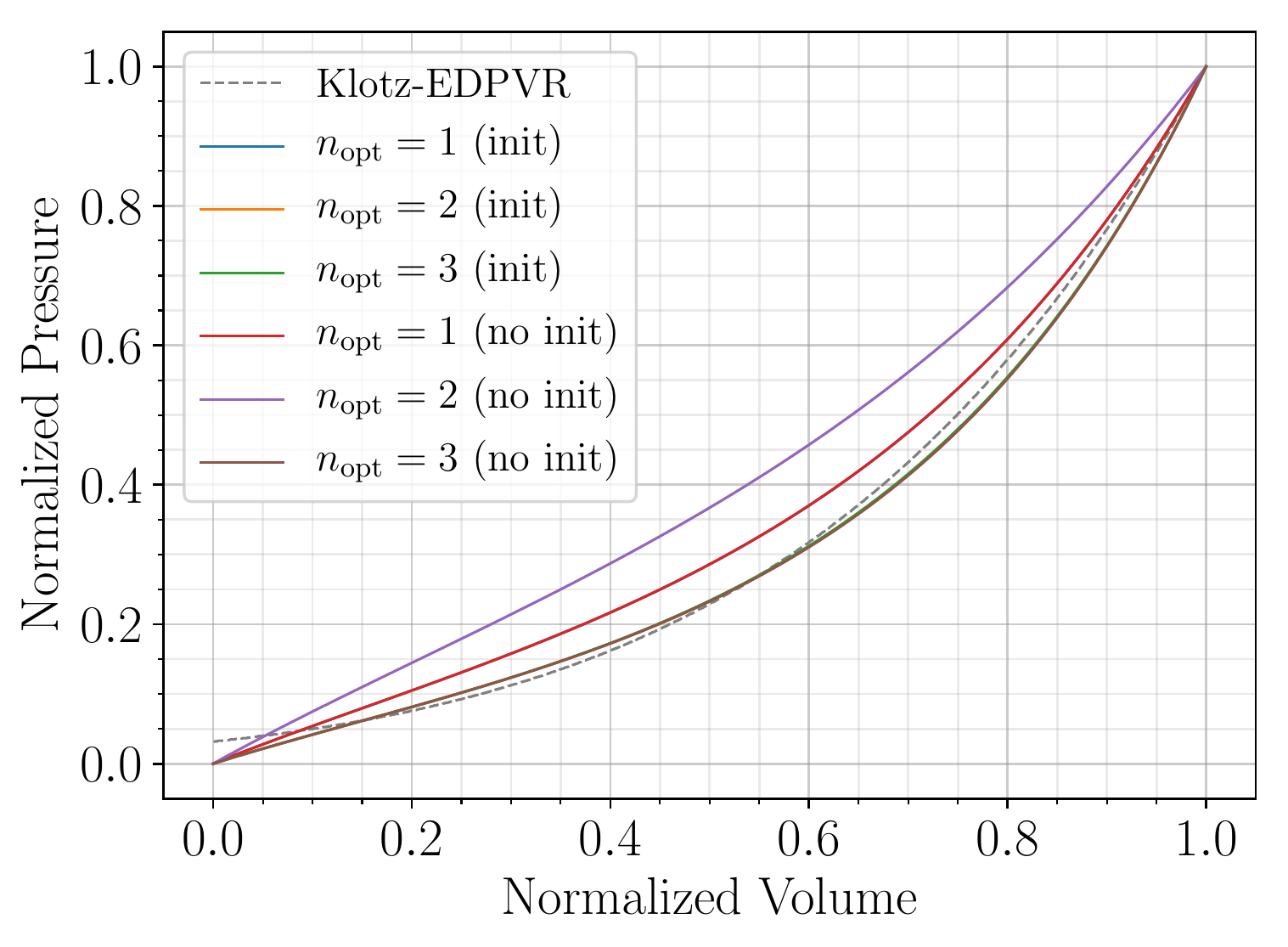}
  %\caption{Case 06-CoA (best fit)}
  %\label{fig:mffall_B62_norm}
\end{subfigure}
\begin{subfigure}{.5\textwidth}
  \centering
  % include second image
  \includegraphics[width=.8\linewidth]{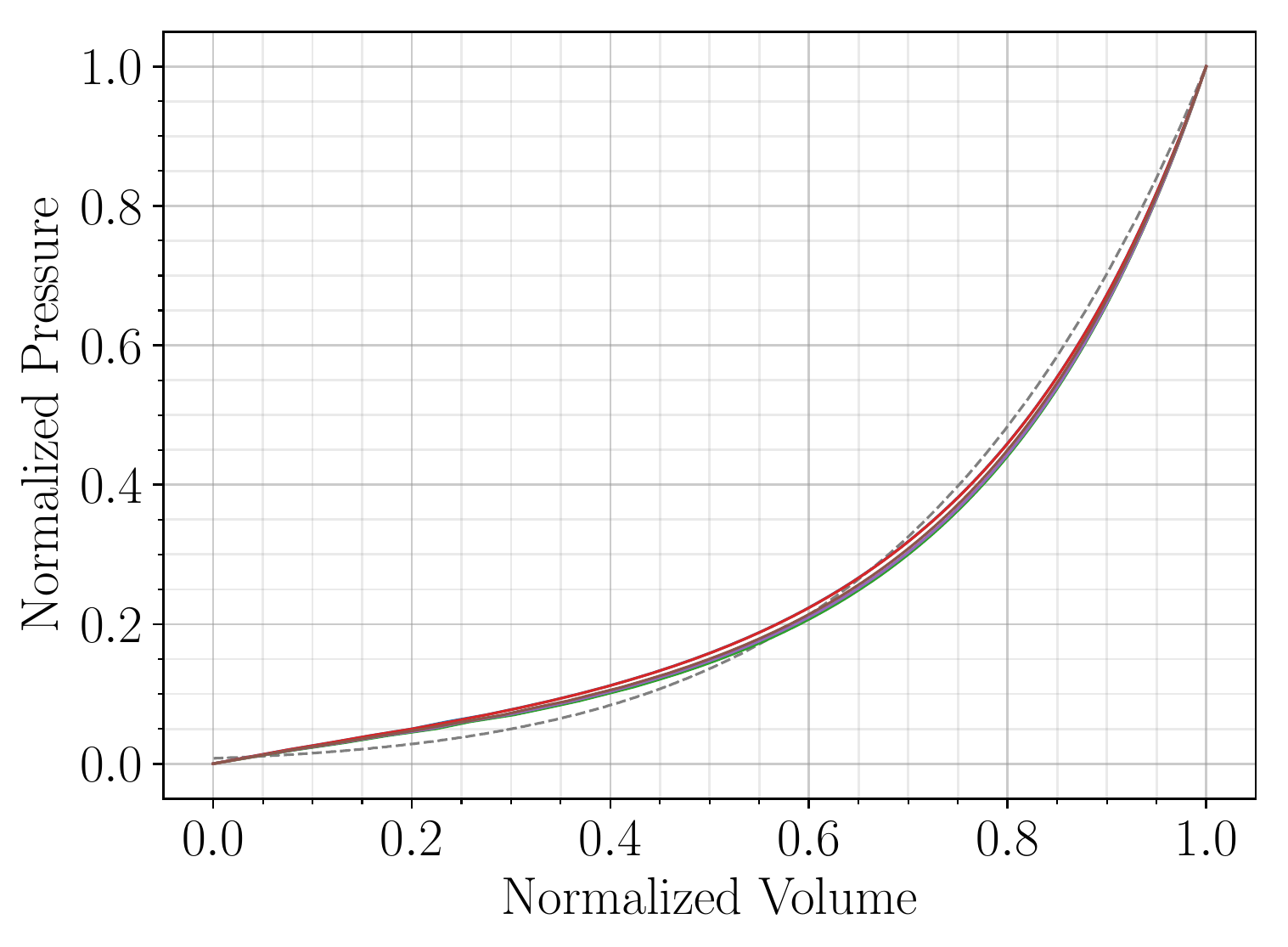}
  %\caption{Case 03-AS (worst fit)}
  %\label{fig:mffall_B82_norm}
\end{subfigure}

\caption{Normalized fitting results of case 06-CoA (left) and case 03-AS (right) executed using the reduced HO constitutive law with
initial guesses from literature (no init) and with a MFF initialization step (init) as initial guess.
Outcomes of the CFF method, optimizing one, two, or three parameters, $n_\mathrm{opt}=\{1,2,3\}$, are compared.}
\label{fig:cff}
\end{figure}

Results are summarized in~\Cref{tab:cff} and in~\Cref{fig:cff} and we see that the worst fit was obtained for case 06-CoA optimizing two parameters and using initial values from the literature (no init),
where the Nelder--Mead method got stuck in a local minumum.
For $n_\mathrm{opt}=1$ similar results were reached when using values from the literature (no init) and initial scalings generated by a MFF (init).
Best fits were acquired for $n_\mathrm{opt}=2$ with MFF initialization and $n_\mathrm{opt}={3}$ in both variants.
For case 03-AS obtained fitted parameters and goodness of fit were very similar for both initialization variants.
Overall, results did not deviate much when one, two, or three parameters were optimized.
In general, simulations run with default parameters from the literature needed more Nelder--Mead iterations compared to simulations with MFF iterations and thus also run-times were longer (\num{>1.5}\;times as long).

% B62
% n_opt=1: less than half as many iterations, shape error twice as big and cost more than three times as big as for n_opt={2,3} and also An bigger. Similar results in terms of shape error, cost and goodness of fit when using init and no init. (except for noinit n_opt=2 --> rerun)
% For n_opt=1 a_scale also very similar.
% For n_opt=3 a_scale=fa_scale similar, but b_scale and fb_scale differ

\begin{table}[htbp]
\begin{center}
\resizebox{\linewidth}{!}{
    \begin{tabular}{lccccccccccccc}
\toprule
& & \multicolumn{6}{c}{Case 06-CoA} & \multicolumn{6}{c}{Case 03-AS} \\
\cmidrule(lr){3-8}\cmidrule(lr){9-14}
 & & & \multicolumn{3}{c}{Fitted Parameters}& \multicolumn{2}{c}{Goodness of Fit} & & \multicolumn{3}{c}{Fitted Parameters}& \multicolumn{2}{c}{Goodness of Fit}\\
\cmidrule(lr){4-6}\cmidrule(lr){7-8}\cmidrule(lr){10-12}\cmidrule(lr){13-14}
Mode & $n_\mathrm{opt}$ &  it. &
      $a_{\mathrm{scale}}$ &  $b^{\mathrm{iso}}_{\mathrm{scale}}$ & $b^{\mathrm{aniso}}_{\mathrm{scale}}$ & $r^{V_0,\mathrm{rel}}$  & $r^{A_n,\mathrm{rel}}$ &
 it. & $a_{\mathrm{scale}}$ &  $b^{\mathrm{iso}}_{\mathrm{scale}}$ & $b^{\mathrm{aniso}}_{\mathrm{scale}}$ & $r^{V_0,\mathrm{rel}}$  & $r^{A_n,\mathrm{rel}}$  \\
&  & & & & & [$\%V_\mathrm{ed}^\mathrm{dat}$] & [$\% A_\mathrm{klotz}$] & & & & & [$\%V_\mathrm{ed}^\mathrm{dat}$] & [$\% A_\mathrm{klotz}$]\\

\midrule
init & 1  & 17 & $0.3475$ & $0.5000$  & $0.5000$ & $0.027$ & $11.13$ & 9 & $0.2982$ & $0.5000$   & $0.5000$ & $0.024$ & $14.72$\\
     & 2  & 46 & $0.2365$ & $0.6493$  & $0.6493$ & $0.002$ & $8.30$  & 73 & $0.2588$ & $0.5252$   & $0.5252$ & $0.001$ & $15.43$\\
     & 3  & 94 & $0.2371$ & $0.6827$  & $0.6274$ & $0.001$ & $8.20$  & 78 & $0.2571$ & $0.5524$   & $0.5144$ & $0.002$ & $15.59$\\
\midrule
no init & 1 & 20 & $0.3477$ & $0.5000$  & $0.5000$ & $0.025$ & $11.15$ & 22 & $0.2969$ & $0.5000$   & $0.5000$ & $0.016$ & $14.72$\\
        & 2 & 41 & $0.5427$ & $0.3255$  & $0.3255$ & $0.009$ & $28.13$ & 80 & $0.2587$ & $0.5252$   & $0.5252$ & $0.000$ & $15.40$\\
        & 3 & 130& $0.2382$ & $0.7321$  & $0.5948$ & $0.008$ & $8.43$  & 167& $0.2588$ & $0.5979$   & $0.4881$ & $0.001$ & $15.03$\\
\bottomrule
\end{tabular}%
}
\caption{Fitting results in terms of fitted scaling parameters and measures of goodness of fit for the two example cases executed
using the reduced HO constitutive law with parameters from literature \cite{Holzapfel2009Constitutive} (no init) as initial guess
and with an MFF initialization step (init).
Outcomes of the CFF method, optimizing one, two, or three parameters, $n_\mathrm{opt}=\{1,2,3\}$, are listed with number of iterations (it.) of the Nelder--Mead algorithm.}
\label{tab:cff}
\end{center}
\end{table}%

\begin{figure}[htbp]
\begin{subfigure}{.5\textwidth}
  \centering
  % include first image
  \includegraphics[width=.8\linewidth]{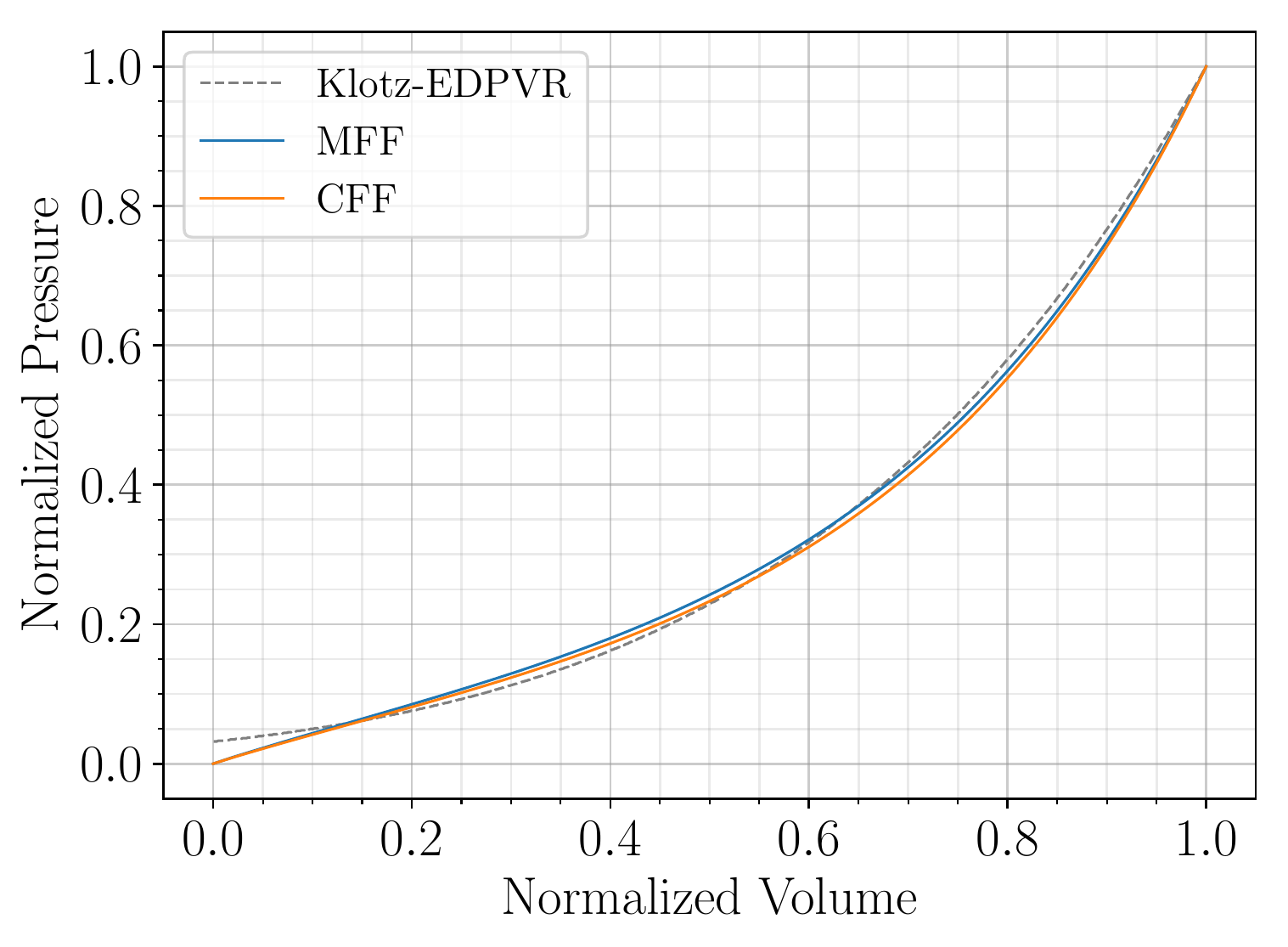}
  %\caption{Case 06-CoA}
  %\label{fig:mffall_B62_norm}
\end{subfigure}
\begin{subfigure}{.5\textwidth}
  \centering
  % include second image
  \includegraphics[width=.8\linewidth]{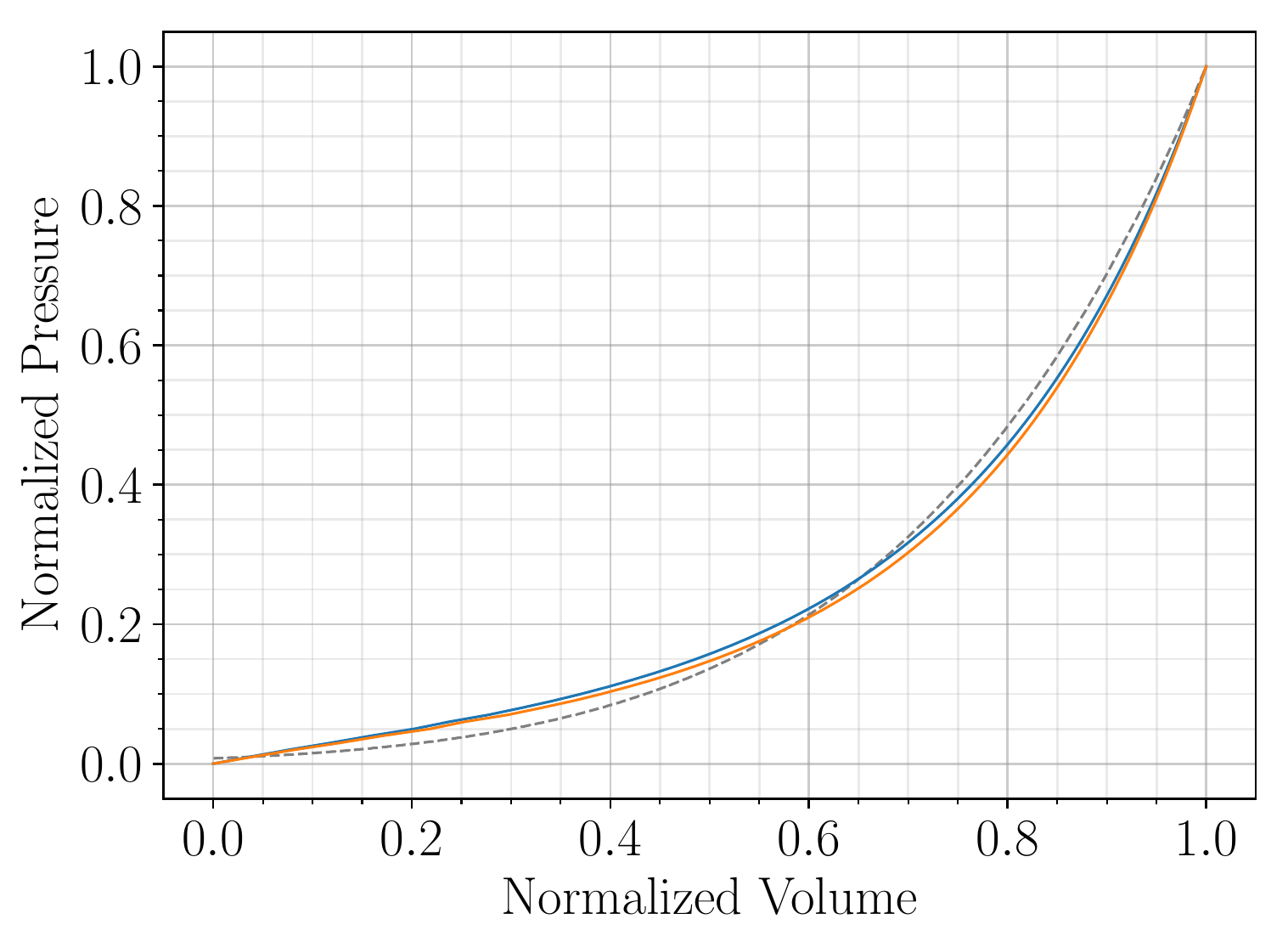}
  %\caption{Case 03-AS}
  %\label{fig:mffall_B82_norm}
\end{subfigure}

\caption{Comparison of fitting outcomes of MFF and CFF for case 06-CoA (left) and case 03-AS (right).}
\label{fig:app-mffvscff}
\end{figure}

\paragraph{Comparison of MFF and CFF approaches} \label{app:mffvscff}
Fitting outcomes of MFF (\Cref{tab:mff-allcases}) and CFF (\Cref{tab:cff}) agree well in terms of fitted parameters and measures of goodness of fit.
Also the PV-curves of the MFF and CFF approach match almost exactly in \Cref{fig:app-mffvscff} where the MFF result is compared to the best fit of the CFF method.
However, since a full unloading step has to be solved in each Nelder--Mead iteration, the number of passive inflation simulations for the CFF approach is significantly larger compared to the MFF method:
$79/156/516$ vs 8 for 06-CoA and 110/460/835 vs $10$ for 03-AS for CFF no init and $n_\mathrm{opt}=1/2/3$ fitted parameters vs. MFF;
thus also the run-time of the CFF approach was considerably longer compared to the MFF approach:
330/546/2095\;min vs \SI{32}{\min} for 06-CoA and 541/2197/3974\;min vs \SI{61}{\min} for 03-AS for CFF no-init
and $n_\mathrm{opt}=1/2/3$ fitted parameters vs. MFF.

\FloatBarrier
%
%-----------------------------------------------------------------------------------------------------------
\section{Discussion}
\subsection{Comparison to other methods and the state-of-the-art}
This study describes a novel methodology to fit passive mechanical parameters of the myocardium. Additionally, the algorithm computes a stress-free
unloaded reference configuration where a new fail-safe feature was implemented that improves the robustness of a previously introduced
backward-displacement method.
We have demonstrated that the presented MFF method is an efficient, robust, and versatile approach for the automated identification of model parameters from
clinical image data using the empirical EDPVR as proposed by~\citet{Klotz2007Computational} as a fitting target.
With this, the MFF algorithm only requires image or mesh data from one time instance
during diastolic filling and a single measured data point of the EDPVR as inputs.
Hence, the MFF method does not depend on a known approximation of the reference configuration
which is a limitation of previous works~\cite{Hadjicharalambous2015analysis,Nasopoulou2017improved}.
This is of particular interest as often only a single anatomical snapshot
taken within the diastatic window is available from clinical studies.
Alternatively, previously generated geometries can be used for the MFF algorithm and
thus no additional segmentation work is required to estimate material parameters for already existing meshes.

Contrary to other studies that jointly estimate material parameters and the unloaded reference
configuration~\cite{Asner2016estimation,Finsberg2018efficient,Finsberg2018estimating,Nikou2016effects},
we did not require to fix any parameters of the material law.
To the best of our knowledge this is thus the first study that allows to fit not only for the predicted
volume of the reference configuration $V_0$ but also for the shape of the EDPVR.
In opposition to studies above, we could apply our algorithm to much finer meshes (up to \num{1.5} million elements) and
still keep simulations tractable with regard to computational cost.
This is essential for many translational application of computational models
in industry or in the clinic where time constraints apply.

Parameter estimation results are in line with previous studies
where experimental parameters from the literature were also often considered
as too stiff~\cite{Kallhovd2019,sack2018construction,wang2013structure}.
Fittings are in a similar range compared to~\cite[Figure~10]{Hadjicharalambous2015analysis}
and~\cite[Table~2]{Nikou2016effects}
for the Guccione law; and~\cite[Table~6]{Asner2016estimation} and~\cite[Table~A1]{Finsberg2018efficient}
for the one-fiber HO law;
in these studies the authors were also able to reproduce the Klotz curve.
Comparison of parameter values presented in other studies is difficult
as these did not take the Klotz relation into account~\cite{Nasopoulou2017improved, Xi2013estimation}
or used different ratios between parameters~\cite{sack2018construction}.

To verify our novel MFF approach we implemented an \emph{ad hoc} CFF based optimization procedure.
Here, MFF and CFF approaches yield similar fitting results distinguishing the MFF approach as the superior method due to significantly shorter run-times
and a higher robustness. While the MFF approach converged for all examined patient-cases, the more expensive CFF approach did not.
This was due to time constraints on clusters and numerical problems stemming from instabilities of the forward solver
owing to very soft material parameters chosen by the Nelder--Mead algorithm.
For the latter case, we returned a very high cost, however, for some patient-specific geometries this still led to convergence problems.
We further showed that even if a CFF approach is pursued, the MFF method can be used as a fast approach to generate an initial guess to improve convergence and efficiency.

Finally, compared to some above-mentioned studies, the proposed method can be coupled with relative ease to established FE software packages.
It only requires a limited number of passive inflation simulations, a standard experiment in all continuum
mechanics simulators,
and basic least-squares fitting tools which are available through open-source packages such as SciPy.
\subsection{Computational costs}
For all patient-specific cases, run-times for the MFF method (between \SI{10.85}{\min} and \SI{74.18}{\min}) were only 2 to 3 times higher compared to a standard passive-inflation experiment.
Note that a further speed-up is possible by using only one Newton step during unloading, making the MFF method sometimes even faster than a single passive inflation experiment.
However, in this case we observed stair-casing effects in the simulated PV-curve that might affect results to a minor degree.
Also, a considerably lower number of loading steps will accelerate simulation times. In this scenario, the least-squares fitting to the piece-wise linear PV-curve could be less accurate.

The high efficiency of the parameter fitting is of utmost importance as the computational burden imposed by high-resolution patient-specific models
demands fast numerical methods to keep simulations tractable.
\subsection{Versatility of the workflow with respect to material laws and FE formulations}
The MFF approach works for various material models including the most widely used passive cardiac tissue models, e.g.,
HO-type and Fung-type materials as introduced in~\cite{Holzapfel2009Constitutive,Gultekin2016,Demiray1972Note,Guccione1995Finite,Usyk2000Effect},
see~\Cref{sec:diff-matlaws}.
Run-times and iteration numbers of the MFF algorithm varied to a small degree with complexity and anisotropy of the model.
Convergence with criteria discussed in~\Cref{sec:error_estimates} was reached for all constitutive laws within \num{10} iterations.
The MFF did not work for the isotropic neo-Hookean and Mooney--Rivlin materials sometimes used for cardiac mechanics~\cite{chabiniok2012,saintemarie2006}.
With these constitutive laws the Klotz-curve could not be reproduced and a similar non-physiological PV-response was observed,
as also reported in~\cite[Figure 12]{Hadjicharalambous2015analysis}.
This is suggestive of such materials being inappropriate
to model cardiac tissue.

We could further show that the MFF approach can be used with different FE formulations such as P1-P0 elements and locking-free stabilized P1-P1 and MINI elements.
Here, the locking-free elements resulted in material parameters that correspond to a stiffer material as the
fitting compensates for locking effects in the P1-P0-element formulation.
As expected, run-times are significantly higher for locking-free elements;
still the efficiency of the MFF approach
allowed for unloading and parameter estimation within reasonable time frames.
%The difference of prestress generated with P1-P0 and locking-free elements and its effect on electromechanical
%simulations is beyond the scope of this work and will be addressed in a future study.
%
\subsection{Goodness of fit and sensitivity analysis}\label{sec:disc:sensitivity}
We were able to show that our method allows to match the predicted stress-free volume $V_0^\mathrm{klotz}$ and the given end-diastolic volume
$V_\mathrm{ed}^\mathrm{dat}$ almost perfectly while reproducing the shape of the PV-curve within uncertainty of the empiric Klotz relation
(\Cref{tab:mff-allcases}).
It is important to note that the EDPVR cannot be accurately determined \emph{in vivo} for a number of fundamental limitations. First, the LV volume during diastolic filling depends on the pressure difference between LV cavity and intra-pericardial pressure
and is also influenced by the presence of time-varying non-zero active stresses.
Time-varying intra-pericardial pressure is non-negligible and influenced by intra-thoracic pressure modulated by breathing and, as such,
cannot be recorded easily beyond specifically designed
experiments~\cite{holt1960,holt1970}.
Further, tails of the active stress transients during diastole depend on
cytosolic calcium, the state of Troponin-C buffering as well as the spatially varying sarcomere geometries. None of these factors can be monitored.
As such, the time course of active stresses during diastole must be considered unknown, albeit efforts have been invested to estimate these \cite{Xi2014}.
The Klotz relation has been shown to be robust across species
and a range of pathological conditions under \emph{ex vivo} conditions
where both volume and intra-cavitary pressure can be controlled with higher accuracy,
but, owing to its empirical nature, must be considered an approximation
of the EDPVR in an individual patient under \emph{in vivo} conditions.

Independently of the uncertainties a Klotz-based surrogate EDPVR is afflicted with,
the application to the patient cohort resulted in excellent fittings to the Klotz curve
for all cases with a goodness of fit measure $r^{A_n,\mathrm{rel}}$
ranging between \SI{7.38}{\%} and \SI{14.77}{\%}.
Also $r^{V_0,\mathrm{rel}}$ was very low for all cases, although a small trend towards better fits in the CoA cases was observed.
Note that only for these cases the measured pressure was available, see also~\Cref{sec:sa-edp}.

The influence of the input pressure $p_\mathrm{ed}^\mathrm{dat}$ was noticeable, resulting in different
shapes of the Klotz EDPVR (\Cref{fig:sa-edp-klotz,fig:sa-fibers})
and therefore in different material parameters.
This influenced the goodness of fit to some extent, hence,
the sensitivity to this measure should be considered,
especially in the case
when invasively measured pressures are not available.

Fiber orientations are known to have a great impact on cardiac mechanics simulations~\cite{Rodriguez-Cantano2019},
thus, the influence of variations of the fiber angle $\alpha$ on fitting performance is an integral point of
the sensitivity analysis.
Variations of fiber angles resulted in very similar outcomes in terms of goodness of fit as well as computational cost.
The altered stiffness due to altered fiber directions was compensated by the variation of material parameters, hence resulting in similar PV-curves.

Changes in initial guesses for model parameters were not reflected in the fitting outcome at all and resulted in the same fitted material parameters and curves.
Only the run-times for unloading varied when using initial guesses
lying further from the final result.
This robustness of the approach is essential for the interpretation of material parameters for
characterizing patient pathology and understanding changes in material properties under HF conditions.

The fitting outcome was not sensitive to variations of the bulk modulus $\kappa$ when using locking-free elements. However, for P1-P0-elements the shape error increased for higher values of $\kappa$,
compensating for the increased locking of the P1-P0-elements. As in all simulations modeling nearly-incompressible behavior of soft tissue,
special care has to be taken when setting this parameter for
simple linear elements.

% P1-P0: r_shape (ml): k=650:9.57, k=1000:10.86, k=3000:17.12, k=5000:22.46
% P1-P1: r_shape (ml) < 6.25 for all values of k

Finally, the choice of the model function~\eqref{eq:model_function} was motivated by the mathematical structure of the
constitutive laws and the Klotz relation.
In~\ref{sec:app-res-modelfunc}, we could demonstrate that several other model functions
can be used for the fitting, all leading to excellent results.
As the MFF approach is applicable to many kinds of target loading curves other than the Klotz relation,
the definite choice of the model function is specific to the problem.

\subsection{Limitations}
First, the workflow presented in this paper relies on the empiric Klotz relation.
However, the approach also works in the exact same fashion for other experiment-based PV-curves,
discrete data points obtained from loading experiments,
or experimental/clinical \emph{in vivo} measurements of the EDPVR.

Second, as residual strain is not considered, in general, this method does not generate a uniform fiber-stretch field which is generally assumed
at end-diastolic state~\cite{carruth2016,fung2013biomechanics_circulation,omens1990residual}.
Heterogeneity in fiber stretch in an end-diastolic state
impairs the Frank-Starling mechanism, as shown in a recent modeling study
\cite{Augustin2020CircAdapt}.
In future studies, the presented workflow will be extended
by using ideas based on growth and remodeling \cite{Genet2015heterogeneous,Wu2020reconstructing} to address this issue.

Third, we presented the methodology only for single-chamber LV models, while solid- and electromechanical whole-organ simulations of the heart are becoming feasible~\cite{Augustin2016anatomically,Baillargeon2014,sack2018construction, Strocchi2020publicly}.
Nevertheless, the workflow was successfully tested for bi-ventricular models as well as 4-chamber models using one homogeneous material for the whole heart and patient-specific pressures in the different chambers.
We did not present these simulations as an empirical estimation for the EDPVR was only available for the LV given as the Klotz law.
To the best of our knowledge, no comparable approximation of EDPVR
exists for the right ventricle or the atria.

Finally, we cannot provide a rigorous proof that the resulting unloaded configuration and material parameter sets are unique.
However, we could show that the method was robust with respect to initial values leading to the same result for all simulations, see~\Cref{sec:disc:sensitivity}.
Further, as we are fitting a strictly convex function~\eqref{eq:model_function}, we can assume that the Levenberg--Marquardt algorithm
results in a unique parameter set~\cite{Boyd2004}.
Since the MFF method requires that the estimated parameters are close to this solution, see~\Cref{eq:err-param},
we can expect that the MFF method is robust with regard to material parameters.
%
%-----------------------------------------------------------------------------------------------------------
\section{Conclusion}
We report on a novel MFF approach combining the fitting passive mechanical parameters of the myocardium to the shape of the EDPVR
and the simultaneous estimation of the unloaded reference configuration.
The algorithm only requires image or mesh data from one time instance during diastolic filling and a single measured PV data point of the EDPVR as inputs.
The MFF is efficient, robust, and versatile and can be applied to reproduce clinically-relevant PV relationships
for patient-specific LV anatomical models
within clinically easily feasible time frames in a fully automated fashion.
Thus, the method constitutes a further step forward
towards a realistic representation of LV passive mechanical function.
As such, the MFF is a pivotal building block in workflows
for building computational digital twin models of human cardiac EM function at scale,
to facilitate the generation of virtual cohorts in translational applications.
%
%-----------------------------------------------------------------------------------------------------------
\section*{Acknowledgments}
This research has received funding from the European Union's Horizon 2020 research and innovation programme under
the ERA-NET co-fund action No.~680969 (ERA-CVD SICVALVES) funded by the Austrian Science Fund (FWF), Grant~I~4652-B to CMA\@.
Additionally, the project was supported by BioTechMed-Graz, Austria as part of the BioTechMed-Graz Flagship Project ILearnHeart to GP\@.
Simulations for this study were performed on the Vienna Scientific Cluster (VSC-4),
which is maintained by the VSC Research Center in collaboration with the Information Technology Solutions of TU Vienna.

\pagebreak
\appendix
\hypertarget{app}
{\section{Klotz relation}}
\label{app:klotz}
Considering the limited availability of clinical data of the EDPVR, the computational method proposed
by~\citet{Klotz2007Computational} is utilized, which enables prediction of the EDPVR by a
single measured PV-pair. According to this seminal work the stress-free volume $V_0^\mathrm{klotz}$ of the LV can be  empirically determined by
\begin{equation}
    \label{eq:const-v0}
    V_0^\mathrm{klotz} = {V}_\mathrm{ed}^\mathrm{dat}
                          \left(0.6 - 0.006{p}_\mathrm{ed}^\mathrm{dat}\right),
\end{equation}
where ${V}_\mathrm{ed}^\mathrm{dat}$ and ${p}_\mathrm{ed}^\mathrm{dat}$ is a measured PV-pair at end-diastole.
Further, the EDPVR is described by the power law
\begin{equation}
   \label{eq:const-powerlaw}
    p = \alpha {V}^{\beta},
\end{equation}
where $p$ is the cavity pressure in \si{\mmHg},
$V$ is the cavity volume in \si{\milli\litre},
and the constants $\alpha$ and $\beta$ are defined by the relations
\begin{equation}
    \label{eq:const-alpha}
    \alpha = \frac{30}{\left(V_{30}^\mathrm{klotz}\right)^\beta} \qquad \text{ and } \qquad
    \beta = \frac{\log\left(\sfrac{p_\mathrm{ed}^\mathrm{dat}}{30}\right)}{\log\left(\sfrac{V_\mathrm{ed}^\mathrm{dat}}{V_{30}^\mathrm{klotz}}\right)}.
\end{equation}
Here, $V_{30}^\mathrm{klotz}$ is the estimated cavity volume at a pressure of \SI{30}{\mmHg}, given by
\begin{equation}
    \label{eq:const-v30}
    {V}_{30}^\mathrm{klotz} = {V}_{0}^\mathrm{klotz} + \frac{V_\mathrm{ed}^\mathrm{dat} - {V}_{0}^\mathrm{klotz}}
    {\left(\sfrac{p_\mathrm{ed}^\mathrm{dat}}{A_\mathrm{n}}\right)^{1/B_\mathrm{n}}},
\end{equation}
where ${A}_\mathrm{n}$ and ${B}_\mathrm{n}$ were determined empirically as \SI{27.78}{\mmHg} and
\num{2.76} respectively.~\Cref{eq:const-alpha} requires that ${p}_\mathrm{ed}^\mathrm{dat} \leq$ \SI{22}{\mmHg} (\SI{2.93}{\kPa}), which applies to all patient cases in the CARDIOPROOF cohort (see~\Cref{tab:mff-allcases}).

%To avoid a singularity in Equation~\eqref{eq:const-alpha}, $\alpha$ and $\beta$
%were redefined as
%
%\begin{equation}
%    \label{eq:const-alpha2}
%   \alpha = \frac{p_\mathrm{ed}^\mathrm{dat}}{\left(V_\mathrm{ed}^\mathrm{dat}\right)^\beta}
%   \qquad \text{ and } \qquad \beta =
%       \frac{\log\left(\sfrac{p_\mathrm{ed}^\mathrm{dat}}{15}\right)}
%            {\log\left(\sfrac{V_\mathrm{ed}^\mathrm{dat}}{V_{15}}\right)}
%\end{equation}
%
%for ${p}^\mathrm{ed}^\mathrm{dat}$ greater than \SI{22}{\mmHg}. $V_{15}$ is determined analytically as
%
%\begin{equation}
%   \label{eq:const-v15}
%    V_{15} = 0.8 (V_{30} - V_0) + V_0.
%\end{equation}

\section{Choice and comparison of model functions}
\label{sec:app-res-modelfunc}
\begin{figure}[htbp]
\centering
%\begin{adjustbox}{minipage=\linewidth, scale=0.86}
    \begin{subfigure}{.45\textwidth}
        \centering
        % include first image
        \includegraphics[width=1.0\linewidth]{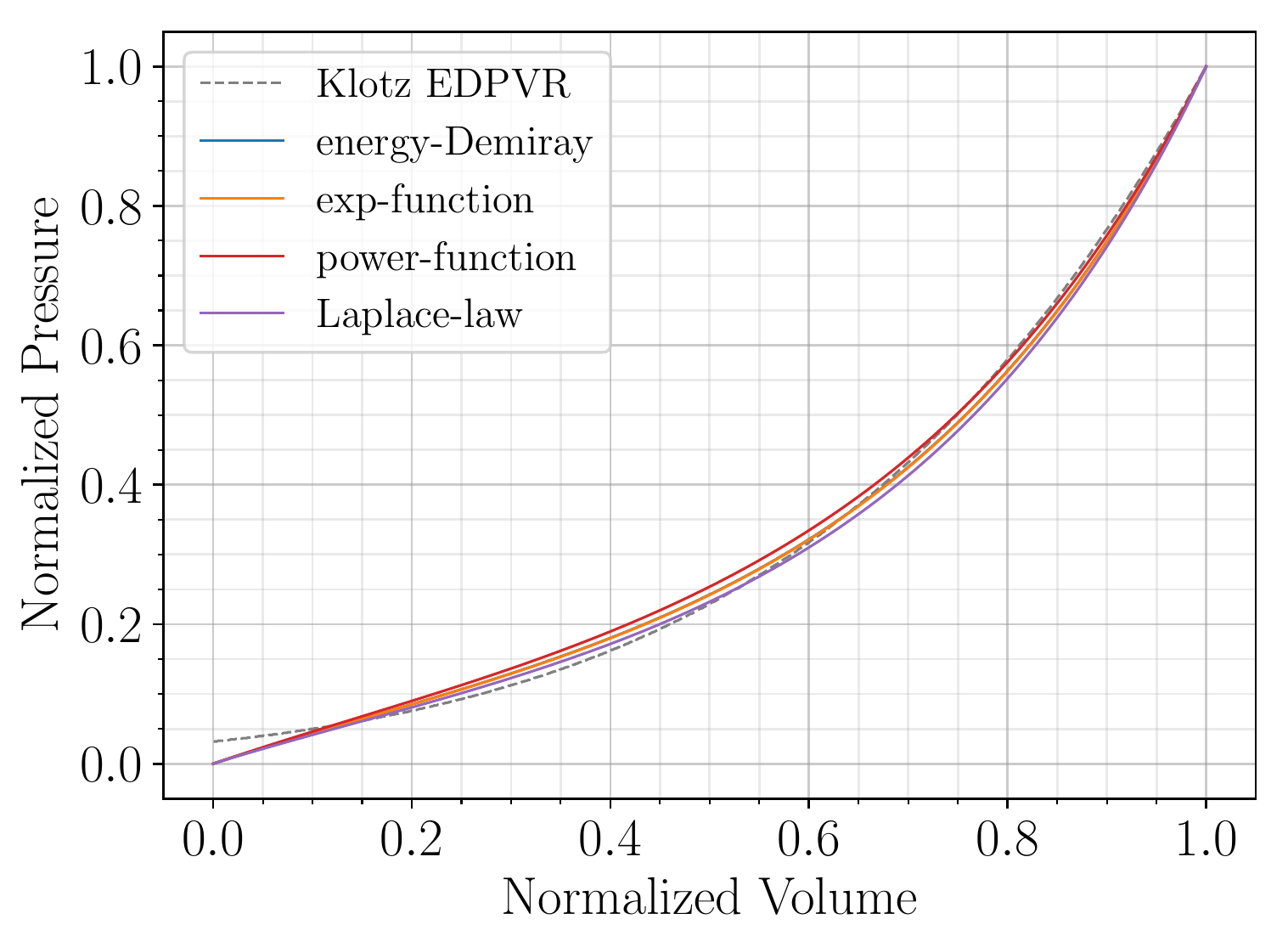}
        %\caption{Case 06-CoA}
        %\label{fig:best}
    \end{subfigure}
    \begin{subfigure}{.45\textwidth}
        \centering
        % include second image
        \includegraphics[width=1.0\linewidth]{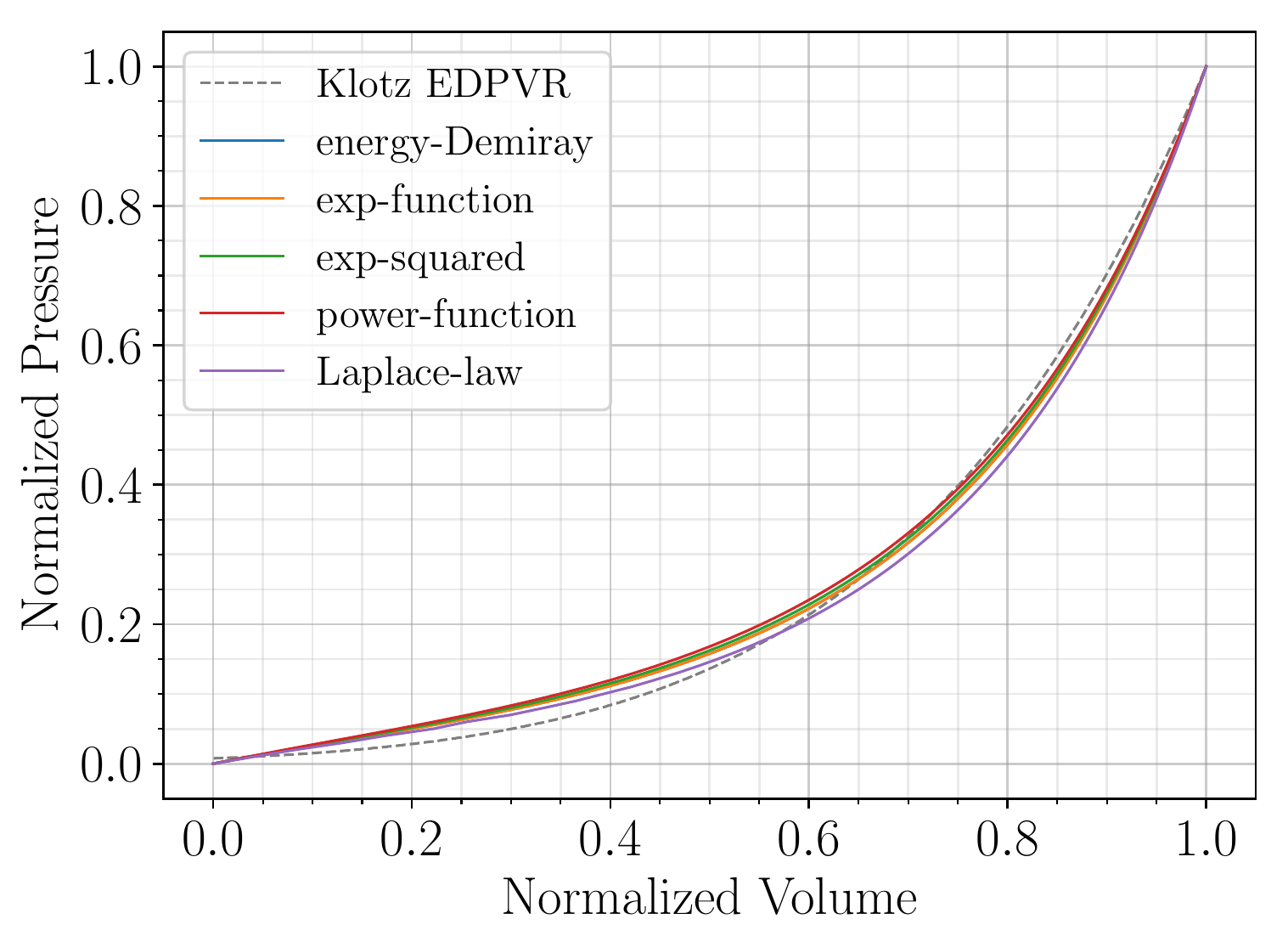}
        %\caption{Case 03-AS}
         %\label{fig:worst}
    \end{subfigure}
%\end{adjustbox}

 \caption{Fitting results of case 06-CoA (left) and 03-AS (right) using different model functions are shown. The normalized Klotz EDPVR (dashed gray) is visualized along with the respective fitted curves.}
  \label{fig:mff-modelfunctions}
\end{figure}

As default we chose the model function in~\Cref{eq:model_function} for its similarity to the constitutive laws and to keep $b\neq 0$.
Nevertheless, the fitting method also worked for other model functions, e.g., those listed below:
\begin{flalign*}
 && \Phi_1(x, x_0)&=\frac{a}{2b}\left\{
  \exp\left[b\left(\frac{x-x_0}{x_0}\right)\right]-1\right\},&\hspace{-4em}\text{(energy-Demiray)}  \\
 && \Phi_2(x, x_0)&=a\left\{
  \exp\left[b\left(\frac{x-x_0}{x_0}\right)\right]-1\right\}, &\hspace{-4em}\text{(exp-function)} \\
 && \Phi_3(x, x_0)&=\frac{a}{2b}\left\{
  \exp\left[b\left(\frac{x-x_0}{x_0}\right)^2\right]-1\right\}, &\hspace{-4em}\text{(exp-squared)}\\
 && \Phi_4(x, x_0)&=a\left(\frac{x-x_0}{x_0}\right)^b, &\hspace{-4em}\text{(power-function)}\\
 && \Phi_5(x, x_0)&=a\left(\frac{x-x_0}{x_0}\right)\exp\left[b\left(\frac{x-x_0}{x_0}\right)^2\right]
  \left[\left(\frac{x+v_\mathrm{wall}}{x}\right)^{\sfrac{2}{3}}-1\right],&\text{(Laplace-law)}
\end{flalign*}
with parameters $a$ and $b$.
Similar to~\Cref{sec:model_function_fitting}, a Levenberg--Marquardt least-squares algorithm was used (i) to fit the model function $\Phi_i(x, x_0)$
to the Klotz relation, with $x$ the volumes as predicted by the Klotz power law~\eqref{eq:const-powerlaw} and
$x_0$ the stress-free volume $V_0^\mathrm{klotz}$~\eqref{eq:const-v0}; and
(ii) to fit the model function $\Phi_i(x, x_0)$ to the re-loading curve in the current step $k$ of the unloading algorithm, with
$x$ the volumes at the different loading points and $x_0$ the cavitary volume of the current reference configuration $\bm{X}^k$.
All model functions are designed such that $\Phi(x_0, x_0)=0$ to ensure loading curves that have zero pressure at the stress-free reference configuration with volume $V_0$.
The first model function (energy-Demiray), the same as in~\Cref{eq:model_function}, was inspired by the constitutive law of Demiray~\eqref{eq:Demiray};
the second (exp-function) is a standard exponential fitting function;
the third (exp-squared) was inspired by the anisotropic contributions in the HO models \eqref{eq:HolzapfelOgdenMyocardialModel};
the fourth (power-function) is related to the Klotz power law~\eqref{eq:const-powerlaw}; and
the fifth (Laplace-law) was inspired by an extension of the Laplace law~\cite{mirsky1973assessment}
to take the volume of the LV wall, $v_\mathrm{wall}$, into account:
\begin{align*}
    \boldsymbol{\sigma} = \frac{p}{\left(\frac{x + v_\mathrm{wall}}{x}\right)^{\sfrac{2}{3}}-1},
\end{align*}
where the stress tensor $\boldsymbol{\sigma}$ at pressure $p$ is computed from the
constitutive law (exp-squared) above
\begin{align*}
    \boldsymbol{\sigma} = \frac{\partial \Phi_3(x,x_0)}{\partial \lambda},  \quad
    \lambda:=\left(\frac{x-x_0}{x_0}\right).
\end{align*}
Here, $x$ is a substitute for the volume of the cavity at pressure $p$ for the Klotz law and the reloading, respectively;
$x_0$ a substitute for the reference volume; and $\lambda$ a strain-like value.
\paragraph{Results} The unloading and parameter estimation was performed as in~\Cref{sec:results_all_cases} for cases 06-CoA and 03-AS.
Normalized fitting results for all different model functions and both cases are shown in~\Cref{fig:mff-modelfunctions} and~\Cref{tab:mff-model}.
We see that all model functions work well for case 03-AS with only minor differences in the goodness of fit and the fitted parameters.
For case 06-CoA the fitting with the (exp-squared) model did not converge and the fitting with the power-function
showed a considerably slower convergence compared to the other model functions. Differences between (energy-Demiray) and (exp-function) are
very small, both show fast convergence and excellent fitting results, rendering these model functions a favorable choice for the fitting.
We noticed for all cases, also visible in~\Cref{fig:mff-modelfunctions}, that the fitting with the (Laplace-law) model function gives results that
are closest to the Klotz curve in the lower pressure range but further afar in the higher pressure range, overall resulting in the
largest values of the area error $r^{A_n,\mathrm{rel}}$. However, the (Laplace-law) fitting always had the lowest deflection error
defined as
\begin{equation}
\label{eq:err-max-deflection}
    r^{\text{shape}}=\max\limits_{p\in[0,p_\mathrm{ed}]}
      \left| V^\mathrm{klotz}(p)-V^\mathrm{sim}(p)\right|.
\end{equation}
Hence, for certain cases it can be a good alternative to the (energy-Demiray) function.

Model functions $\Psi_i$ given above are only a small subset of functions that we tried for our fitting to the Klotz curve.
All of them showed satisfying results and convergence rates but other functions might work as well for the procedure.
Since the Klotz curve resembles an exponential function, it is not surprising that the (energy-Demiray) and (exp-function) worked best.
However, as our method would work for all kinds of target EDPVR other than the Klotz law,
the choice of the model function is specific to the problem.
% run-times
% B62-CoA: t_ul= 31.5-100.98  (similiar run-times except power-functionn 3 times as long, t_val ~ 10 min)
% B82-AS: t_ul= 37.48 -139.65 (similiar run-times except power-functionn 3.7 times as long, t_val ~ 13 min)

\begin{table}[htbp]
\resizebox{\linewidth}{!}{
    \begin{tabular}{lcccccccc}
\toprule
            & \multicolumn{4}{c}{Case 06-CoA} & \multicolumn{4}{c}{Case 03-AS} \\
\cmidrule(lr){2-5}\cmidrule(lr){6-9}
   & \multicolumn{2}{c}{Fitted Parameters} & \multicolumn{2}{c}{Goodness of Fit}& \multicolumn{2}{c}{Fitted Parameters} & \multicolumn{2}{c}{Goodness of Fit} \\
\cmidrule(lr){2-3}\cmidrule(lr){4-5}\cmidrule(lr){6-7}\cmidrule(lr){8-9}
Model Function &  $a_{\mathrm{scale}}$ &  $b_{\mathrm{scale}}$  & $r^{V_0,\mathrm{rel}}$  & $r^{A_n,\mathrm{rel}}$  &  $a_{\mathrm{scale}}$ &  $b_{\mathrm{scale}}$  & $r^{V_0,\mathrm{rel}}$  & $r^{A_n,\mathrm{rel}}$ \\
  & & & [$\%V_\mathrm{ed}^\mathrm{dat}$] & [$\%A_\mathrm{klotz}$]& & & [$\%V_\mathrm{ed}^\mathrm{dat}$] & [$\%A_\mathrm{klotz}$] \\

\midrule
energy-Demiray      & $0.2576$ & $0.6257$ & $0.13$ & $7.39$ & $0.2971$ & $0.5076$ & $0.19$ & $14.77$ \\
exp-function        & $0.2558$ & $0.6267$ & $0.11$ & $7.36$ & $0.2968$ & $0.5078$ & $0.19$ & $14.77$ \\
exp-squared         & -        & -        & -      & -      & $0.3120$ & $0.4956$ & $0.16$ & $14.64$ \\
power-function      & $0.2878$ & $0.6076$ & $0.59$ & $7.14$ & $0.3260$ & $0.4721$ & $0.19$ & $14.59$ \\
Laplace-law         & $0.2358$ & $0.6587$ & $0.28$ & $8.40$ & $0.2605$ & $0.5391$ & $0.25$ & $15.55$ \\
\bottomrule
\end{tabular}%
}
\caption{Fitting results for case 06-CoA and 03-AS using different model functions are shown in terms of fitted scaling parameters and measures of goodness of fit.}
\label{tab:mff-model}%
\end{table}%
%
%%%%%%%%%%%%%%%%%%%%%%%%%%%%%%%%%%%%%%%%%%%%%%%%%%%%%%%%%%%%%%%%%%%%%%%%%%%%%%%
% BIBLIOGRAPHY
\bibliographystyle{elsarticle-num-names}
\bibliography{bibliography}

\end{document}